\documentclass[lettersize,journal,onecolumn,draftclsnofoot]{IEEEtran}
\usepackage{cite,comment}
\usepackage{amsmath,amssymb,amsfonts, mathtools}
\usepackage{algorithmic}
\usepackage{graphicx}
\usepackage{makecell}
\usepackage{textcomp}
\usepackage{dsfont}
\def\BibTeX{{\rm B\kern-.05em{\sc i\kern-.025em b}\kern-.08em
    T\kern-.1667em\lower.7ex\hbox{E}\kern-.125emX}}

\def \OO {{O}}
\def \oo {{o}}
\newcommand{\pr}{\mathbb{P}}

\newcommand{\N}{\mathbb{N}}

\newcommand{\limit}{\underset{n \rightarrow \infty}\lim}
 
\newcommand{\comp}{^{\rm c}}

\newtheorem{theorem}{\bf Theorem}[section]
\newtheorem{definition}[theorem]{Definition}
\newtheorem{lemma}[theorem]{Lemma}

\newcommand{\hh}{\mathbb{H}(n;K)} 
\newcommand{\hhn}{\mathbb{H}(n;K_n)} 
\newcommand{\hhdn}{\mathbb{H}(n;K_n,\gamma_n)} 
\newcommand{\nodes}{\mathcal{N}}

\allowdisplaybreaks

\begin{document}
\title{On the Robustness, Connectivity and Giant Component Size of Random  K-out Graphs }
\author{Eray Can Elumar, \IEEEmembership{Student Member, IEEE}, Mansi Sood, \IEEEmembership{Student Member, IEEE}, and Osman Ya\u{g}an, \IEEEmembership{Senior Member, IEEE}
\thanks{Parts of the material was presented at the 2021 IEEE International Symposium on Information Theory (ISIT) \cite{can_isit21}, and the 2023 IEEE International Conference on Communications (ICC) \cite{can_icc2023}. This work was supported in part by the  Office of Naval Research (ONR) through Grant  N00014-21-1-2547, by the CyLab IoT Initiative, and by the National Science Foundation through Grant CCF-1617934.}
\thanks{Eray Can Elumar is with the Department of Electrical and Computer Engineering, Carnegie Mellon University, Pittsburgh, PA 15213 USA (e-mail: eelumar@andrew.cmu.edu). }
\thanks{Mansi Sood is with the Department of Electrical and Computer Engineering, Carnegie Mellon University, Pittsburgh, PA 15213 USA (e-mail: msood@andrew.cmu.edu).}
\thanks{Osman Ya\u{g}an is with the Department of Electrical and Computer Engineering and CyLab, Carnegie Mellon University, Pittsburgh, PA 15213 USA (e-mail: oyagan@andrew.cmu.edu).} }

\maketitle

\begin{abstract}
Random K-out graphs are garnering interest in designing distributed systems including secure sensor networks, anonymous crypto-currency networks, and differentially-private decentralized learning. In these security-critical applications, it is important to model and analyze the resilience of the network to node failures and adversarial captures. Motivated by this, we analyze how the connectivity properties of random K-out graphs vary with the network parameters $K$, the number of nodes ($n$), and the number of nodes that get failed or compromised ($\gamma_n$). In particular, we study the conditions for achieving  \emph{connectivity} {with high probability} and for the existence of a \emph{giant component}  with formal guarantees on the size of the largest connected component in terms of the parameters $n,~K$, and $\gamma_n$. Next, we analyze the property of \emph{$r$-robustness} which is a stronger property than connectivity and leads to resilient consensus in the presence of malicious nodes. We derive conditions on $K$ and $n$ under which the random K-out graph achieves r-robustness with high probability. We also provide extensive numerical simulations and compare our results on random K-out graphs with known results on Erd\H{o}s-R\'enyi (ER) graphs.

%
\end{abstract}


\section{Introduction}
 \subsection{Motivation and Background}
 \label{sec:introduction}


In recent years, the rapid proliferation of affordable sensing and computing devices has led to rapid growth in technologies powered by the IoT (Internet of Things). A key challenge in this space is to develop network models for generating a securely connected ad-hoc network in a distributed fashion while minimizing operational costs.    

With its unique connectivity properties, a class of random graph models known as the \emph{random $K$-out graphs} has found many applications in the design of ad-hoc networks. A random K-out graph \cite{Bollobas,FennerFrieze1982,Yagan2013Pairwise}, denoted as $\hh$, is an undirected graph with $n$ nodes where each node forms an edge with $K$ distinct nodes chosen  uniformly at random. Random K-out graphs are known to achieve connectivity easily, i.e., with far fewer edges $(O(n))$ as compared to classical random graph models including Erd\H{o}s-R\'enyi (ER)  graphs \cite{erdos61conn,Bollobas}, random geometric graphs \cite{penrose2003random}, and random key graphs \cite{yagan2012zero}, which all require $O(n \log n)$ edges for connectivity.  In particular, it is known \cite{Yagan2013Pairwise, FennerFrieze1982} that random K-out graphs are connected with high probability ({\em whp}) when $K \geq 2$. This had led to their deployment in several applications including  random key predistribution schemes for secure communication in sensor networks \cite{mansi_pairwise1,mansi_pairwise2,yagan2012modeling},  differentially-private distributed
averaging algorithms \cite{2020dprivacy}, 
anonymity preserving cryptocurrency networks \cite{FantiDandelion2018}, and distributed secure mapping of network addresses in next-generation internet architectures \cite{SIAM2021paper}.

In the context of sensor networks, random K-out graphs have been used \cite{Yagan2013Pairwise, yagan2012modeling, yavuz2015designing} to  analyze  the performance of the random {\em pairwise} key predistribution scheme and its heterogeneous variants \cite{eletreby2020connectivity,sood2020size}. The random {\em pairwise} scheme works as follows. Before deployment, each sensor chooses $K$ others uniformly at random. A unique {\em pairwise} key is given to each node pair where at least one of them selects the other. After deployment, two sensors can securely communicate if they share a pairwise key. The topology of the sensor network can thus be represented by a random K-out graph; each edge of the random K-out represents a secure communication link between two sensors.
Consequently, random K-out graphs have been analyzed  to answer key questions on the values of the parameters $n$ and $K$ needed to achieve certain desired properties, including connectivity at the time of deployment \cite{FennerFrieze1982,Yagan2013Pairwise}, connectivity under {\em link} removals  \cite{yagan2012modeling,yavuz2015designing}, and unassailability \cite{yagan2016wireless}.  Despite many prior works on random K-out graphs, very little is known about its connectivity properties when some of its {\em nodes} are removed. This is an increasingly relevant problem since many IoT networks are deployed in remote and hostile environments where nodes may be captured by an adversary, or fail due to harsh conditions.

Another application of random K-out graphs is in distributed learning, where a key goal is to perform  computations on user data without compromising the privacy of the users. Random K-out graphs have  recently been used to construct the communication graph in a differentially-private federated averaging scheme called the GOPA (GOssip Noise for Private Averaging) protocol \cite[Algorithm~1]{2020dprivacy}.  
According to the GOPA protocol, a random K-out graph is constructed on a set of nodes, of which an {\em unknown} subset is {\em dishonest}. 
It was shown in \cite[Theorem~3]{2020dprivacy} that the privacy-utility trade-offs achieved by the GOPA protocol are tightly dependent on the subgraph on {\em honest} nodes being {\em connected}. When the subgraph on honest nodes is not connected, it was shown that the performance of GOPA is tied to the {\em size} of the connected components of  honest nodes.  Since dishonest users can be modeled as randomly deleted nodes, analyzing the connectivity and giant component size of 
random K-out graphs under node deletions is the key in understanding the performance of the GOPA protocol. 


 \subsection{Properties of Interest for Random K-out Graphs in Distributed Computing Applications}

In the context of  applications discussed in the previous section, we can  identify several key properties of random K-out graphs that need to be well-understood for performance evaluation and efficient design of the underlying systems. We believe that the graph properties discussed here can also be useful in facilitating new applications of random K-out graphs in different  fields, 
akin  to  our  recent  work  \cite{sood2021tight}  paving the way to new applications by establishing connectivity guarantees in the finite node regime.

A key metric in quantifying the utility of a network is \emph{connectivity} which is defined as the existence of a path of edges between every node pair \cite{graph_theory_book}. Connectivity ensures that all agents in the network can communicate with one another and no node is isolated from the network.  In practice, resource constraints limit the number of links that can be established in the network. Thus, a key goal is to design a  \emph{resiliently} connected network while keeping the number of links to be established within operational constraints. Depending on the resource constraints and mission requirements of the application at hand, it may suffice to ensure a weaker notion of connectivity or in cases where agents may routinely fail  or compromise, we may even need a stronger notion of connectivity. 

In resource-constrained environments, preserving connectivity despite  node failures may not be feasible and the goal instead might be to ensure that there is a {\em large enough}, connected sub-network of users, also known as a \emph{giant} component.
For example, it may suffice to aggregate the temperature readings of the majority of sensors deployed in a field to get an estimate of the true temperature. Another example is the power grid network where it is essential to ensure supply to the majority of the users in the event of failures.


In addition to ensuring that the network remains \emph{resiliently} connected in the event of node failures, it is often desirable to ensure that \emph{consensus} can be achieved even in the presence of adversarial agents. In \cite{sundaram_robust}, it was shown that network connectivity is not sufficient to characterize consensus when nodes use a certain class of local filtering rules. In particular, it was shown that consensus can be reached in graphs that have the property of being sufficiently \emph{robust}. This is formally quantified by the property of \emph{$r$-robustness}, which was  introduced in   \cite{sundaram_robust}. A graph is said to be \emph{$r$-robust} if, for every disjoint subset pair that partitions the graph, at least one node in one of these subsets is adjacent to at least $r$ nodes in the other set.

The $r$-robustness property is especially useful in applications of {\em consensus dynamics}, where  parameters of several agents get aligned after a sufficiently long period of local interactions. In another example, it was shown  \cite{robust_consensus} that if the network is $(2F + 1)$-robust (for some non-negative integer $F$), then the nodes in the network can reach consensus even when there are up to $F$ malicious nodes in the neighborhood of every correctly-behaving node. Thus,  $r$-robustness 
is particularly important for applications based on consensus dynamics in adversarial environments.  Moreover, $r$-robustness is known \cite{robust_consensus} to be a stronger property than $r$-connectivity and thus can provide guarantees on the connectivity of the graph when up to $r-1$ nodes in the graph are removed. Due to this importance, $r$-robustness has been studied in various graph models in the random graph literature, such as the ER graph and the Barab\'asi-Albert model \cite{zhang2012robustness}. $r$-robustness has been studied in \cite{can_cdc2021} in our prior work; and the results presented here improve upon those results by providing a sharper threshold for $r$-robustness.  

 \subsection{Main Contributions}


With these motivations in mind, this paper aims to fill the gaps in the literature on the connectivity and robustness properties of random K-out graphs. We provide a comprehensive set of results on the connectivity and size of the giant component of the random K-out graph when some of its nodes are {\em dishonest}, have {\em failed}, or have been {\em captured}. We further analyze the conditions required for ensuring $r$-robustness of the random K-out  graph. Our main contributions are summarized below:
\begin{enumerate}
    \item Let $\hhdn$ denote the random graph obtained after removing $\gamma_n$ nodes, selected uniformly at random, from the random K-out graph $\hhn$. We provide a set of conditions for $K_n$, $n$, and $\gamma_n$ under which $\hhdn$ is connected {\em with high probability} (whp). This is done for both cases where $\gamma_n=\Omega(n)$ and $\gamma_n=o(n)$, respectively. 
 Our result for $\gamma_n = \Omega(n)$ (see Theorem \ref{theorem:thmc_1}) significantly improves a prior result \cite{YAGAN2013493} on the same problem and leads to a {\em sharp} zero-one law for the connectivity of the random K-out graph under node deletions. Our result for the case $\gamma_n = o({n})$ (see Theorem \ref{theorem:thmc_2}) expands the existing threshold of $K_n \geq 2$ required for connectivity by showing that the graph is  connected whp for $K_n \geq 2$ even when $o(\sqrt{n})$ nodes are deleted.
    \item We derive conditions on $K_n$, $n$, $\gamma_n$ that lead to a {\em giant component} in $\hhdn$ whp and provide an upper bound on the number of nodes not contained in the giant component. This is also done for both cases $\gamma_n=\Omega(n)$ and $\gamma_n=o(n)$; see Theorem~\ref{theorem:thmg_3} and Theorem~\ref{theorem:thmg_5}, respectively.  An important consequence of this result is to establish $K_n \geq 2$ as a sufficient condition to ensure {\emph whp} the existence of a giant component in the random K-out graph despite the removal of $o(n)$ nodes in the network.    
    \item Using a novel proof technique, we show that $K \geq 2r$ when $r \geq 2$ is sufficient to ensure that
the random K-out graph $\hh$ is $r$-robust {\emph whp} (see Theorem~\ref{theorem:thmr_1}). Since it is already known that $\hh$ is {\em not} $r$-robust {\emph whp} when $K < r$, this result is tight up to at most a multiplicative factor of two (and it is is much tighter than the condition  established in \cite{can_cdc2021}).
\item
We also provide a comparison of  random K-out graphs with ER graphs. We determine that 
    random K-out graphs are much more robust in terms of $r$-robustness property, and also attain connectivity and admit a giant component with fewer edges compared to ER graphs. 
   \item
    Combining our theoretical results with numerical simulations, we highlight the usefulness of random K-out graphs as a topology design tool for efficient design of secure, resilient and robust distributed networks. 

\end{enumerate}

 \subsection{Organization of the Paper}
The rest of this article is organized as follows. In Section \ref{sec:model}, we introduce the notation used across this article and the network model, the random K-out graph, and extend this model to account for node deletions. In Section \ref{sec:Main Results}, we present the main results along with the simulation results and provide a detailed discussion. In Section \ref{sec:proof}, we provide the proof of all Theorems presented in Section  \ref{sec:Main Results}. Conclusions are provided in Section \ref{sec:conclusion}.

\section{Notations and Definitions}
\label{sec:model}

All random variables are defined on the same probability space $(\Omega, {\mathcal{F}}, \mathbb{P})$ and probabilistic statements are given with respect to the probability measure $\mathbb{P}$. The complement of an event $A$  is denoted by $A\comp$. The cardinality of a discrete set $A$ is denoted by $|A|$. The intersection of events $A$ and $B$ is denoted by$A\cap B$.
We refer to any mapping $K:\N_0 \rightarrow \N_0$ as a {\em scaling} if it satisfies the condition $ 2 \leq K_n < n, \quad  n=2, 3, \ldots $. All limits are understood with  $n$ going to infinity. If the probability of an event tends to one as $n\rightarrow \infty$, we say that it  occurs with high probability (whp). The statements $a_n = \oo(b_n)$, $a_n=\omega(b_n)$,  $a_n = \OO(b_n)$, $a_n=\Theta(b_n)$, and $a_n = \Omega(b_n)$, used when comparing the asymptotic behavior of sequences $\{a_n\},\{b_n\}$, have their meaning in the standard Landau notation. The asymptotic equivalence $a_n \sim b_n$ is used to denote the fact that  $\lim_{n \to \infty} \frac{a_n}{b_n}=1$. Finally, we let $\langle d \rangle$  denote the mean node degree of a graph. 

\begin{definition}[Random K-out Graph]
{\sl
The random K-out graph is defined on the vertex set $V:=\{v_1, \ldots, v_n\}$ as follows. Let $\nodes:=\{1,2,\dots,n\}$ denote the set vertex labels. For each $i \in \nodes$,  let $\Gamma_{n,i} \subseteq \nodes \setminus i$ denote the set of $K_n$ labels, selected uniformly at random, corresponding to the nodes selected by $v_i$. It is assumed that  $\Gamma_{n,1}, \ldots , \Gamma_{n,n}$ are mutually independent. Distinct nodes $v_i$ and $v_j$ are adjacent, denoted by $v_i \sim v_j$ if at least one of them picks the other. Namely, 
\begin{align}
v_i \sim v_j ~~\quad \textrm{if} ~~~\quad [j \in \Gamma_{n,i}] ~\vee~ [i \in \Gamma_{n,j}].  \vspace{-1mm}
\label{eq:Adjacency}
\end{align}

The set of neighbors of node $i$ is denoted by $\mathcal{V}_i:=\{j \in \nodes \setminus i: v_i \sim v_j\}$, and the degree of node $i$ is denoted as $d_i = |\mathcal{V}_i|$. The random graph defined on the vertex set $V$ through the adjacency condition \eqref{eq:Adjacency} is called a random K-out graph \cite{frieze2016introduction,Bollobas,Yagan2013Pairwise} 
and is denoted by $\hhn$. 
 }

\label{def:rand_kout}
\end{definition}{}

\begin{definition}[Cut]
\cite[Definition 6.3]{MeiPanconesiRadhakrishnan2008}
For a graph $\mathcal{G}$ defined on the node set $\nodes$, a \emph{cut} is a non-empty subset $S \subset \nodes$ of nodes {\em isolated} from the rest of the graph. Namely,  $S \subset \nodes$ is a cut if there is no   edge between $S$ and $S\comp=\nodes \setminus S$. If $S$ is a cut, then so is $S\comp$.
\label{def:cut}
\end{definition}{}

\begin{definition}[Connected Components]
{\sl
A pair of nodes in a graph $\mathbb{G}$ are said to be {\em connected} if there exists a path of edges connecting them. A {\em connected component} $C_i$ of $\mathbb{G}$ is a subgraph in which any two vertices are connected to each other, and no vertex is connected to a node outside of $C_i$. }

\label{def:concomp}
\end{definition}{}

\begin{definition}[Giant Component]
{\sl
For a graph $\mathbb{G}$ with $n$ nodes, a {\em giant} component exists if its {\em largest} connected component has size $\Omega(n)$. In that case, the largest connected component is referred to as the giant component of the graph. }

\label{def:giantcomp}
\end{definition}{}

\begin{definition}[Connectivity]
A graph $\mathbb{G}$ is {\em connected} if there exists a path of edges between every pair of its vertices.

\label{def:connectivity}
\end{definition}{}

\begin{definition}[$r$-connectivity]
A graph is $r$-connected if it remains connected after the removal of {\em any} set of $r-1$ (or, fewer) nodes or edges.
\label{def:r_connectivity}
\end{definition}{}


\begin{definition}[$r$-reachable Set] \cite[Definition 6]{sundaram_robust}
{\sl
For a graph $\mathbb{G}$ and a subset $S$
of nodes $S \subset \nodes$, we say $S$ is  $r$-reachable  if $\exists $i$ \in S: |\mathcal{V}_i \setminus S| \geq r$, where $r \in \mathbb{Z}^+$. In other words,  $S$ is
an $r$-reachable set if it contains a node that has at least $r$ neighbors outside $S$.
 }

\label{def:r-reachable}
\end{definition}{}

\begin{figure}[!t]
\vspace{-1mm}
\centering
\includegraphics[scale=0.15]{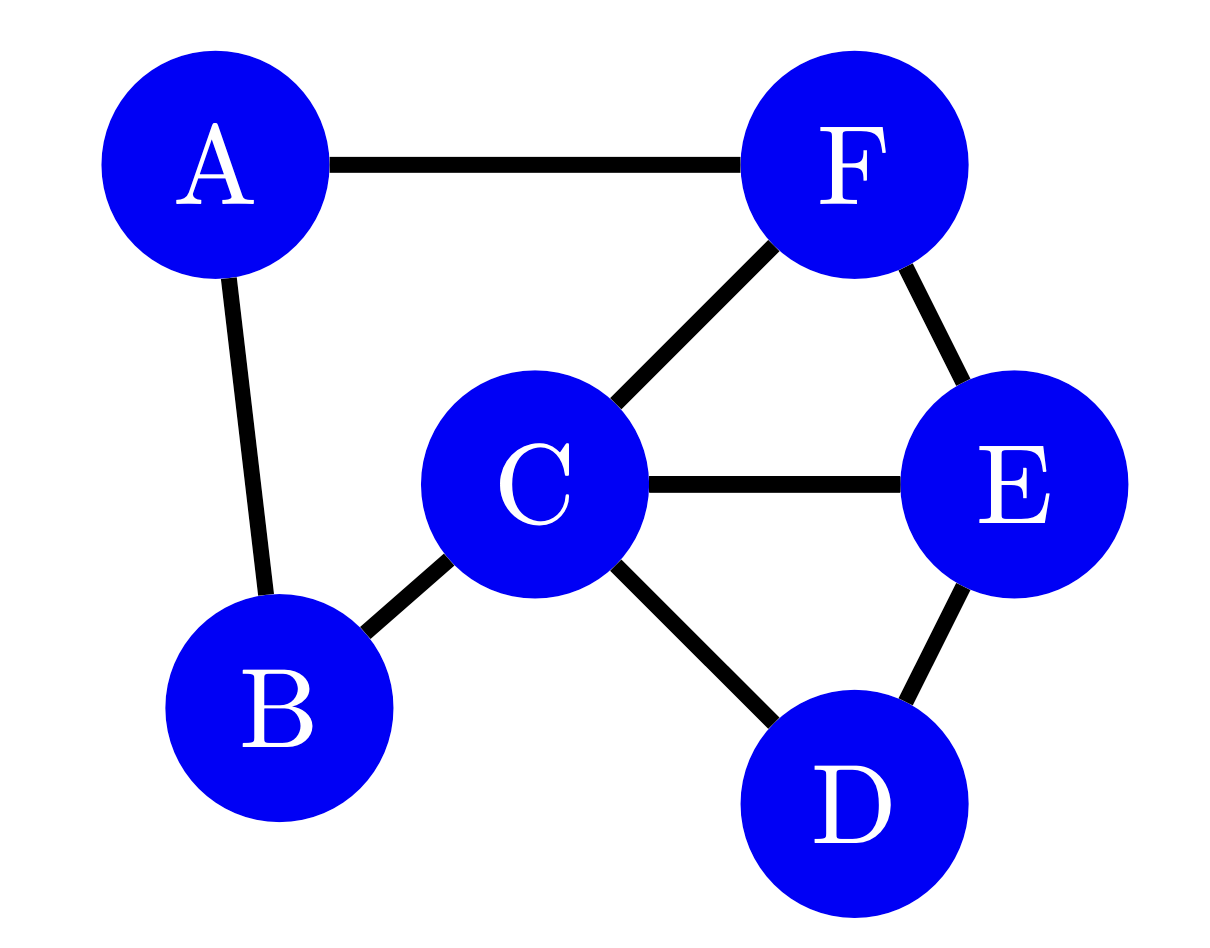}\label{fig:1robust} \vspace{-2mm}
\caption{\sl An example for a 1-robust graph. We see that with the subset pair $S = \{v_A, v_B\}$ and $S^c = \{v_C, v_D, v_E, v_F\}$, both  $v_A$ and  $v_B$ in $S$ have only one neighbor in $S^c$, while  $v_C$ and $v_F $ in $S^c$ have only one neighbor in $S$, meaning both $S$ and $S^c$ are $1$-reachable (but not $2$-reachable). Further, it can be seen that all other subset pairs that partition the graph are also at least $1$-reachable, leading to the graph in Fig.~\ref{fig:fig1rob} being $1$-robust. } 
\label{fig:fig1rob}
\end{figure}

\begin{definition}[$r$-robust Graph] \cite[Definition 6]{sundaram_robust}
{\sl
A graph $\mathbb{G}$ is $r$-robust if for every pair of nonempty, disjoint subsets of $\nodes$, at least one of these subset pairs is $r$-reachable, where $r \in \mathbb{Z}^+$.
 }
 \label{def:r-robust}
\end{definition}{}

It was shown in \cite{robust_consensus} that if a graph is $r$-robust, it is at least $r$-connected. Thus, $r$-robustness is a stronger property than $r$-connectivity. It is also easy to see that when $r=1$, the properties of $r$-robustness and $r$-connectivity are equivalent.

A main goal of this paper is to study the connectivity and giant component size of random K-out graphs  when some of its nodes are {\em failed}, {\em captured}, or {\em dishonest}. To this end, we also consider the following model of random K-out graphs under random removal of nodes. We first let $\gamma_n$ denote the number of removed nodes and assume, for simplicity, that they are selected uniformly at random among all nodes in $V$. 
 Further, we let $D\subset V,\ |D| = \gamma_n$ denote the set of deleted nodes.
 We  then define $\hhdn$ on the vertex set $R = V \setminus D$ and the corresponding set of labels $\nodes_R$, such that  distinct vertices $v_i$ and $v_j$ (both in $R$) are adjacent if they were  adjacent in $\hhn$; i.e., if  $[j \in \Gamma_{n,i}] ~\vee~ [i \in \Gamma_{n,j}]$. For each $i \in \nodes_R$, the set of labels adjacent to node $v_i$ in $\hhdn$ is denoted by $\Gamma_{n-\gamma_n,i} \subseteq \nodes_R \setminus i$.


\section{Main Results}

\label{sec:Main Results}
Our main results are presented in Theorems \ref{theorem:thmc_1}- \ref{theorem:thmr_1} below. Each Theorem addresses a design question as to how the parameter $K_n$ should be chosen to satisfy the desired property on robustness, connectivity or the size of the giant component; see Table \ref{table:compare} for a summary of the main results. The results on connectivity and the size of the giant component are for $\hhdn$, i.e., the random K-out graph when $\gamma_n$ nodes are deleted, while the result
 on $r$-robustness is given for the original graph $\hhn$ (without any node deletion). We provide 
the proofs of all results in Section \ref{sec:proof}.


\subsection{Results on Connectivity}
Let $P(n,K_n,\gamma_n) = \pr \left[ \hhdn \text{ is connected}\right]$.

\begin{theorem}
{\sl 
Let $\gamma_n = \alpha n$ with $\alpha$ in $(0,1)$, and consider a scaling $K:\mathbb{N}_0 \to \mathbb{N}_0$ such that with $c>0$ we
have 
\begin{align}
K_n \sim c \cdot r_1(\alpha, n), \ \  \textrm{where} \quad r_1(\alpha, n) = \frac{\log n}{1 - \alpha - \log \alpha}
\label{eq:threshold_1}
\end{align}
is the threshold function. Then, we have
\begin{align}
& \lim_{n \to \infty} P(n,K_n,\gamma_n) =  \begin{cases}
    1, & \mathrm{if}\quad c > 1\\
    0, & \mathrm{if}\quad 0< c < 1.
  \end{cases}
\label{eq:c_1}
\end{align}
}  \label{theorem:thmc_1}
\end{theorem}
The proof of the {\em one-law} in (\ref{eq:c_1}), i.e., that 
$\lim_{n \to \infty} P(n,K_n,\gamma_n)=1$
if $c>1$, is given in Section \ref{sec:proof}. The {\em zero-law} of (\ref{eq:c_1}),
i.e., that 
$\lim_{n \to \infty} P(n,K_n,\gamma_n)=0$
if $c<1$,
was  established previously in \cite[Corollary 3.3]{YAGAN2013493}. There,
a one-law was also provided: 
under
(\ref{eq:threshold_1}), it was shown that
$\lim_{n \to \infty} P(n,K_n,\gamma_n)$
if $c>\frac{1}{1-\alpha}$, leaving a gap between the thresholds of the zero-law and the one-law. 
Theorem \ref{theorem:thmc_1} presented here fills this gap by establishing a tighter one-law, and constitutes a {\em sharp} zero-one law; e.g., when $\alpha=0.5$, 
 the one-law in \cite{YAGAN2013493} is given with $c>2$, while we show that it suffices to 
have $c>1$. 

\begin{theorem}
{\sl  Consider a scaling $K:\mathbb{N}_0 \to \mathbb{N}_0$. \\
a) If $\gamma_n = o(\sqrt{n})$, then we have
\begin{align}
\lim_{n \to \infty} P(n,K_n,\gamma_n) = 
    1, \quad \mathrm{if}\quad K_n \geq 2 \ \ \forall n
\label{eq:c_2a}
\end{align}
b) If $\gamma_n = \Omega(\sqrt{n})$ and $\gamma_n = o(n)$, and if for some sequence $w_n$, it holds that
$$
K_n = r_2(\gamma_n) + \omega_n, \ \ \textrm{where} \quad 
r_2(\gamma_n) = \frac{\log (\gamma_n)}{\log 2 + 1/2}$$
is the threshold function, then we have
\begin{align}
\lim_{n \to \infty} P(n,K_n,\gamma_n) = 
    1, \quad \mathrm{if}\quad \lim_{n \to \infty}\omega_n = \infty
\label{eq:c_2b}
\end{align}
}  \label{theorem:thmc_2}
\end{theorem}

We remind that random K-out graph is known \cite{FennerFrieze1982,Yagan2013Pairwise} to be connected whp when $K_n \geq 2$. Part $(a)$ of Theorem \ref{theorem:thmc_2} extends this result by showing that having $K_n \geq 2$ is sufficient for the random K-out graph to remain connected whp even when $o(\sqrt{n})$ of its nodes (selected randomly) are deleted. We believe that this result will further facilitate the application of random K-out graphs in a wide range of applications where connectivity despite node failures is crucial.

\subsection{Results on the Size of the Giant Component}

Let $C_{max}(n,K_n,\gamma_n)$ denote the set of nodes in the {\em largest}  connected component  of  $\hhdn$ and let $P_G(n,K_n,\gamma_n,\lambda_n) :=\mathbb{P}[|C_{max}(n,K_n,\gamma_n)| > n - \gamma_n- \lambda_n]$. Namely, 
$P_G(n,K_n,\gamma_n,\lambda_n)$ is the probability that less than $\lambda_n$ nodes are {\em outside} the largest  component of $\hhdn$.

\begin{theorem}
{\sl 
Let $\gamma_n = o(n)$, $\lambda_n = \Omega(\sqrt{n})$ and $\lambda_n \leq \lfloor (n-\gamma_n)/3 \rfloor$. Consider a scaling $K:\mathbb{N}_0 \to \mathbb{N}_0$ and let
$$r_3(\gamma_n,\lambda_n) = 1 +  \frac{\log(1+\gamma_n / \lambda_n)}{\log 2 + 1/2}$$
be the threshold function. Then, we have
\begin{align*}
& \lim_{n \to \infty}  P_G(n,K_n,\gamma_n,\lambda_n) =  1, \quad \mathrm{if}\quad K_n > r_3(\gamma_n, \lambda_n),  \ \ \forall n.
\end{align*}

}  \label{theorem:thmg_3}
\end{theorem}

We remark that  if $\lambda_n=\beta n$ with $0<\beta<1/3$, then  $r_3(\gamma_n,\lambda_n) = 1+o(1)$.
This shows that when $\gamma_n=o(n)$, it suffices to have $K_n \geq 2$ for $\hhdn$ to have a giant component containing $(1-\beta)n$  nodes for arbitrary $0<\beta<1/3$. Put differently, by choosing $K_n \geq 2$, we ensure that even when $\gamma_n=o(n)$ nodes are removed, the rest of the network contains a connected component whose {\em fractional} size is arbitrarily close to 1.


\begin{theorem}
{\sl 
Let $\gamma_n = \alpha n$ with $\alpha$ in  $(0,1)$, and $\lambda_n \leq \lfloor \frac{(1-\alpha)n}{3} \rfloor$. Consider a scaling $K:\mathbb{N}_0 \to \mathbb{N}_0$ and let
$$r_4(\alpha, \lambda_n) = 1 +  \frac{\log(1 + \frac{n \alpha}{\lambda_n}) +\alpha + \log(1-\alpha)}{\frac{1-\alpha}{2} - \log\left(\frac{1+\alpha}{2}\right)}$$
be the threshold function. Then, we have
\begin{align*}
& \lim_{n \to \infty}  P_G(n,K_n,\alpha,\lambda) =  1, \quad \mathrm{if}\quad K_n > r_4(\alpha, x_n), \ \ \forall n.
\end{align*}
}  \label{theorem:thmg_5}
\end{theorem}

It can be seen from this result that $K_n$ needs to scale as $K_n \sim \log(\frac{\alpha n}{\lambda_n })$ for a random K-out graph to have a giant component of size $n-\lambda_n$ when $\alpha n$ of its nodes are removed (or, if each node is independently removed with probability $0<\alpha<1$).  
We also remark that the threshold $r_4(\alpha,\lambda_n)$ is finite when $\lambda_n=\Omega(n)$. This shows that even when a positive fraction of the nodes of the random K-out graph are removed, a finite $K_n$ is still sufficient to have a giant component of size $\Omega(n)$ in the graph. This result can be  useful in applications where it is required to maintain a giant component as efficiently (i.e., with as fewest edges) as possible  even when large scale node failures take place.

\subsection{Result on Robustness}

\begin{theorem}
{\sl 
Define 
$$r^{\star}(K) = \max_{r =1, 2, 3\ldots} \{  \lim_{n \to \infty} P(\hh \text{ is } r \text{-robust}) = \hspace{-0.5mm} 1 \}$$
Then, we have 
$$r^{\star}(K) \geq \lfloor K/2 \rfloor$$
}  \label{theorem:thmr_1}
\end{theorem}

In Theorem \ref{theorem:thmr_1}, we establish a threshold for one-law of $r$-robustness in random K-out graphs, and find that $r^{\star}(K) \geq \lfloor K/2 \rfloor$. In other words, we find that {\em with high probability}, a random K-out graph is $r$-robust when $K \geq 2r, r \geq 2, r \in \mathbb{Z}^+$, and $n \to \infty$. This threshold is much smaller than the previously established threshold \cite{can_cdc2021} of $ K > \frac{2r\left(\log(r) + \log(\log(r) + 1\right)}{\log(2) + 1/2 - \log\left(1+ \frac{\log(2)+1/2}{2\log(r)+5/2+\log(2)}\right) } $ which scales with $r\log r$. Hence, Theorem \ref{theorem:thmr_1} constitutes a sharper one-law for $r$-robustness.

{\small
\begin{table}[t!]
\centering
\tabcolsep=0.1cm
\renewcommand{\arraystretch}{0.9}
\begin{tabular}{ |c|c|c| }
 \hline  
Desired   & Minimum $K_n$ needed to  & Theorem \\
  Property    & achieve the property &  \\ 
 \hline \hline & &  \\
 \makecell{Connectivity   \\ $\gamma_n =  \alpha n$}    &$ (1+\varepsilon) \frac{\log n}{1 - \alpha - \log \alpha}$& Thm. \ref{theorem:thmc_1} 
 \\ 
 \hline
 & &  
 \\ 
 \makecell{Connectivity \\   $\gamma_n = o(\sqrt{n})$}  & $ K_n \geq 2$  &Thm. \ref{theorem:thmc_2}(a)\\
  \hline
 & &  
  \\ 
  \makecell{Connectivity \\  $\gamma_n = w(\sqrt{n})$, $\gamma_n = o(n)$}   & $ \frac{\log (\gamma_n)}{\log 2 + 1/2} + w(1)$  &Thm. \ref{theorem:thmc_2}(b)\\
  \hline
 & & 
  \\  
  \makecell{Giant Component, \\ $\lambda_n=o(n)$, $\gamma_n=o(n)$} & $1 + \frac{\log(1+\frac{\gamma_n}{\lambda_n})}{\log(2) + 1/2}$ &  Thm. \ref{theorem:thmg_3}\\ 
 \hline
 & &  
 \\ 
  \makecell{Giant Component, \\  $\lambda_n=\beta n$, $\gamma_n=o(n)$} & $K_n \geq 2$&  Thm. \ref{theorem:thmg_3}\\ 
  \hline
 & &  
 \\ 
\makecell{Giant Component, \\ $\lambda_n < \lfloor \frac{(1-\alpha)n}{3} \rfloor$, $\gamma_n=\alpha n$}  & $1 + \frac{\log(1 + \frac{\alpha n}{\lambda_n}) + \alpha + \log(1-\alpha)  }{\frac{1-\alpha}{2} - \log(\frac{1+\alpha}{2})}$&Thm. \ref{theorem:thmg_5}\\ 
 \hline
  & &  
 \\ 
   $r$-robustness     & $  K_n \geq 2r$ & Thm. \ref{theorem:thmr_1}  
 \\
 \hline
 
\end{tabular}
\vspace{2mm}
\caption{\sl Summary of our main results providing a condition on $K_n$ needed to achieve a desired property in $\hhdn$ (whp) where $\gamma_n$ denotes the number of deleted nodes.
For the giant component, the desired property is defined as its size to be at least $n-\lambda_n$. In the first row, the result is for any $\varepsilon>0$. On the fifth row, we have $\beta$ in $(0,1/3)$ and on the sixth row, we have $\alpha$ in $(0,1)$. }
 \label{table:compare}
\end{table}
}

\subsection{Discussion}

 In Theorem \ref{theorem:thmc_1}, we improve the results given in \cite{YAGAN2013493}, and with this, we close the gap between the zero law and the one law, and hence establish a sharp zero-one law for connectivity when $\gamma_n= \Omega(n)$ nodes are deleted from $\hhdn$. 
 In Theorem \ref{theorem:thmc_2}, we establish that the graph $\hhdn$ with $\gamma_n = o(n)$ is connected whp when $K_n \sim \log(\gamma_n)$; and when $\gamma_n = o(\sqrt{n})$, $K_n \geq 2$ is sufficient for connectivity. The latter result is especially important, since $K_n \geq 2$ is the previously established threshold for connectivity \cite{FennerFrieze1982}. We improve this result by showing that the graph is still connected with $K_n \geq 2$ even after $o(\sqrt{n})$ nodes (selected randomly) are deleted. 
 
Since most distributed systems require connectivity in the event of node failures, our results can  be useful in many applications of distributed systems, particularly when the resources  on each node is limited and it is critical to achieve desired connectivity and robustness properties using as few edges as possible. For example, in wireless sensor networks, knowing  the minimum conditions needed for connectivity or giant component size  under such failures is crucial as it enables designing them with fewest edges possible per node \cite{yagan2016wireless,di2008redoubtable}, which reduces the communication overhead and potentially the cost of the hardware on each node.

 We also note that Theorems \ref{theorem:thmg_3} - \ref{theorem:thmg_5} constitute the first results concerning the giant component size of random K-out graphs under randomly deleted nodes. In particular, these results help choose the value of $K_n$ for any anticipated level of node failure and  for any given giant component size required, enabling the designs of distributed systems to compromise between efficiency, robustness, and the required giant component size. 
 Thus, we expect these results to be useful in  applications where connectivity is not a stringent condition under node failures, and instead having a certain giant component size is sufficient to continue the operation of the system. 


In Theorem \ref{theorem:thmr_1}, we establish that a random K-out graph is $r$-robust {\em whp} when $K_n \geq 2r$ for any $r \in \mathbb{Z}^+$. This is a much sharper one-law than the previous result  given in \cite{can_cdc2021} where it was shown that $K_n$ needs to scale as  $K_n \sim r \log(r)$ for $r$-robustness. This tighter result  was made possible through several novel steps introduced here. While the proofs in prior work \cite{sundaram_robustness,can_cdc2021} also rely on finding upper bounds on the probability of having at least one subset that is not $r$-reachable, they tend to utilize standard upper bounds for the binomial coefficients ${n \choose k}\leq \left(\frac{en}{k}\right)^k$ and a union bound to establish them. Instead, our proof uses extensively  the Beta function $B(a,b)$ and its properties to obtain tighter upper bounds on such probabilities, which then enables us to establish a much sharper one-law for $r$-robustness of random K-out graphs. We  believe this result will pave the way for further applications of  random K-out graphs in distributed computing applications such as the design of consensus networks in the presence of adversaries.

It is also of interest to compare the threshold of $r$-robustness and $r$-connectivity in random K-out graphs. 
For Erd\H{o}s-R\'enyi graphs, the threshold for $r$-connectivity and $r$-robustness have been shown  \cite{sundaram_robustness} to coincide with each other. For random K-out graphs, we know from \cite{FennerFrieze1982} that $\hhn$ is $r$-connected \emph{whp} whenever $K_n \geq r$, and it is {\em not} $r$-connected \emph{whp} if $K_n < r$.  This leaves a factor of 2 difference between the condition $K_n \geq 2r$ we established for $r$-robustness here and the threshold of $r$-connectivity. Put differently, we know from \cite{FennerFrieze1982} and Theorem \ref{theorem:thmr_1} that for any $r=2, 3, \ldots$
\begin{align}
& \lim_{n \to \infty} \pr \left[ \hhn \text{ is $r$-robust}\right] =  \begin{cases}
    1, & \mathrm{if}\quad K_n \geq 2r\\
    0, & \mathrm{if}\quad K_n< r.
  \end{cases}
\label{eq:osy_new}
\end{align}
For $r=1$, it is instead known that $\lim_{n \to \infty} \pr \left[ \hhn \text{ is $r$-robust}\right]=1$ if only if $K_n \geq 2r=2$. 
Since the currently established conditions for the zero-law and one-law of $r$-robustness are not the same for random K-out graphs (unlike ER graphs where the two thresholds coincide),  there is a question as to whether our threshold of $2r$ is the tightest possible for $r$-robustness. This is currently an open problem and would be an interesting direction for future work, e.g., by establishing a tighter zero-law for $r$-robustness that coincides with the one-law of Theorem \ref{theorem:thmr_1}. Comparison

To put all these results in perspective, we provide comparisons of our results with the results from an Erd\H{o}s-R\'enyi graph $G(n,p)$, which is  one of the most commonly used random graph models. Firstly, in terms of $r$-robustness, it was shown in \cite{sundaram_robustness} that ER graph G(n; p) is $r$-robust \emph{whp} if $p_n = \frac{\log(n) + (r-1) \log(\log(n)) + \omega(1)}{n}$, which translates to a mean node degree of $\langle d \rangle  \sim \log(n) + (r-1) \log(\log(n))$. Since the mean node degree required for random K-out graphs scales as $\langle d \rangle\sim 2 r$, we can conclude that for large $n$, random K-out graphs can be ensured to be $r$-robust \emph{whp} at a mean node degree significantly smaller than the mean node degree required for an ER graph. In terms of connectivity, an ER graph becomes connected whp if $p > \log n / n$ \cite{erdHos1960evolution}, and this translates to having a mean node degree of $\langle d \rangle \sim \log n$. Similarly, when $o(\sqrt{n})$ nodes are removed, the mean node degree required for connectivity scales with $\langle d \rangle \sim \log n$ \cite{erdos61conn}. The $\langle d \rangle$ required for the random K-out graph to be connected whp is much lower, with $\langle d \rangle = O(1)$ when $o(\sqrt{n})$ nodes are removed, and $\langle d \rangle \sim \log(\gamma_n)$ when $\gamma_n=\Omega(\sqrt{n})$ nodes are removed.

 \begin{figure}[!t]
\vspace{-1mm}
\centering
\includegraphics[scale=0.45]{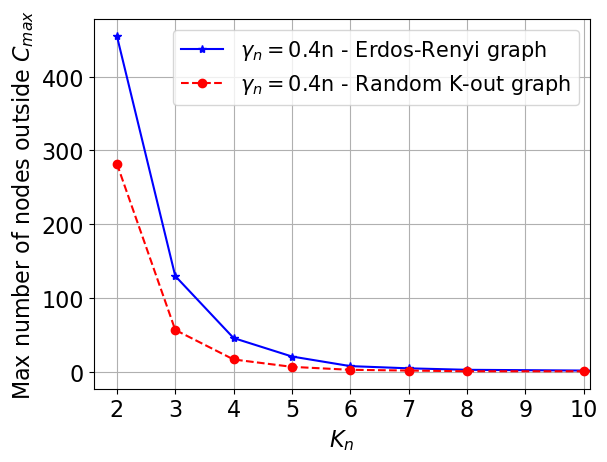}\label{fig:cmp1}
\hspace{0.4mm}
\includegraphics[scale=0.45]{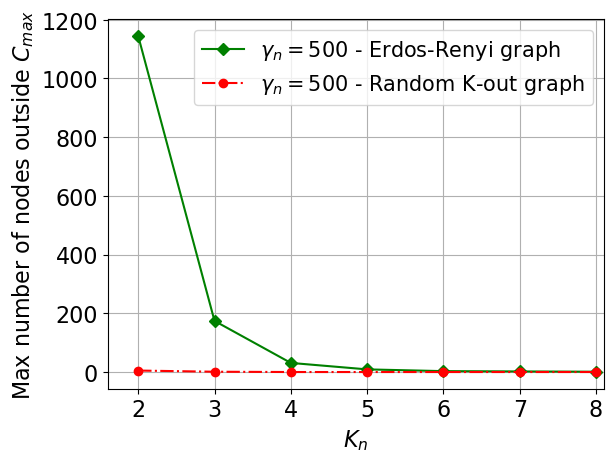}\label{fig:cmp2}  
\caption{\sl Comparison of maximum number of nodes outside the giant component of a random K-out graph $\hhdn$ and an ER graph with same mean node degree when $n = 5000$, $\gamma_n = 0.4n$ (Left); and when $n = 50,000$ and $\gamma_n= 500$ (Right). Each data-point is obtained through 1000 experiments.} 
\label{fig:cmp}
\end{figure}

Next, since we are not aware of any theoretical results on the giant component size of ER graphs under node removals, we compare the size of the giant component in random K-out graphs and ER graphs under node removals through simulations in the finite node regime.  We examine the  maximum number of nodes outside the giant component out of 1000 experiments of a random K-out graph $\hhdn$ and an ER graph $G(n,p)$ with the same mean node degree when $\gamma_n$ nodes are removed from both graphs. 
To ensure that both graphs have the same mean node degree, $p$ in the ER graph is selected as $p = 2K_n/n$. The results are given in Fig.~\ref{fig:cmp} for $n=5000$, $\gamma_n =0.4n$ on (Left), and $n=50,000$, $\gamma_n= 500$ on (Right). As can be seen, the random K-out graph tends to have fewer nodes outside of the giant component than the ER graph and this difference is more pronounced when $\gamma_n$ is smaller. 

In conclusion, we see that when both graphs have the same mean node degree, random K-out graphs are more robust  than ER graphs in terms of connectivity and giant component size under random node removal, and also in terms of the $r$-robustness property. 
This reinforces the efficiency of the K-out construction in many distributed computing applications 
where connectivity in the event of node failures or adversarial capture of nodes is crucial. Similarly, the fact that random K-out graphs tend to achieve $r$-robustness with fewer edges per node than ER graphs (for any  $r=1,2, \ldots$), makes it more suitable in  applications based on distributed consensus. 

\subsection{Simulation Results} \label{sec:simulations}

Since our results are asymptotic in nature, i.e., they have been established in the limit $n \to \infty$, an important question is whether they can also be useful in practical settings where the number $n$ of nodes is finite. 
We check the usefulness to validate Theorems \ref{theorem:thmc_1} - \ref{theorem:thmg_5} under practical settings, 
To answer this,  we examine the probability of connectivity and the number of nodes outside the giant component for the graph $\hhdn$ (random K-out graph with deleted nodes) through computer simulations in two different  setups\footnote{Determining whether a graph is $r$-robust is a co-NP-complete problem  \cite{sundaram_robustness} making it not feasible to check the usefulness of  Theorem \ref{theorem:thmr_1} through computer simulations.}.  

\begin{figure}[!t]
\centering
\includegraphics[scale=0.45]{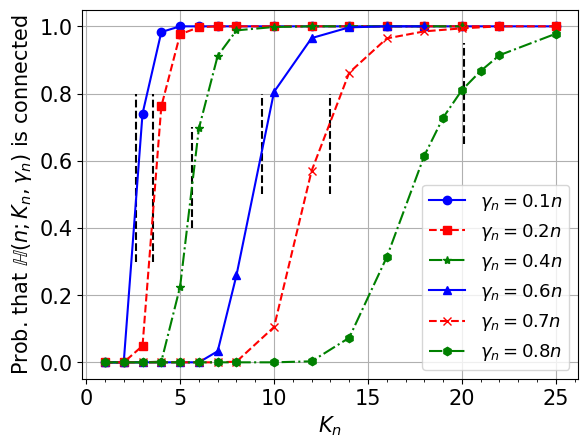}\label{fig:conn1}
\hspace{1mm}
\includegraphics[scale=0.45]{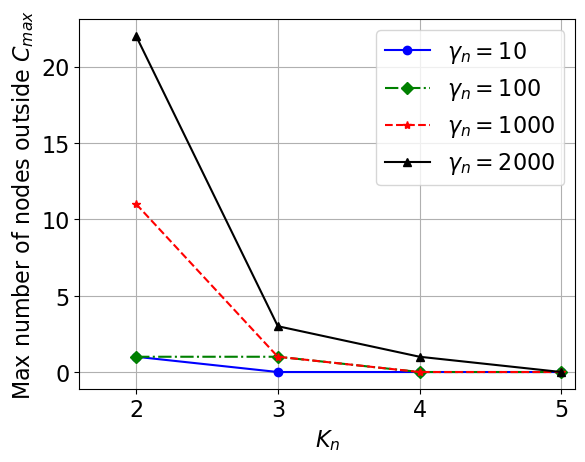}\label{fig:conn2} \vspace{-3mm}
\caption{\sl (Left) Empirical probability that $\hhdn$ is connected for $n = 5000$ calculated from 1000 experiments. The vertical lines are the theoretical thresholds given by Theorem  \ref{theorem:thmc_1}. (Right) Maximum number of nodes outside the giant component of $\hhdn$ for $n = 50,000$ in 1000 experiments. 
\vspace{-4mm}} 
\label{fig:fig1}
\end{figure}

\begin{figure}[!t]
\vspace{-1mm}
\centering
\includegraphics[scale=0.45]{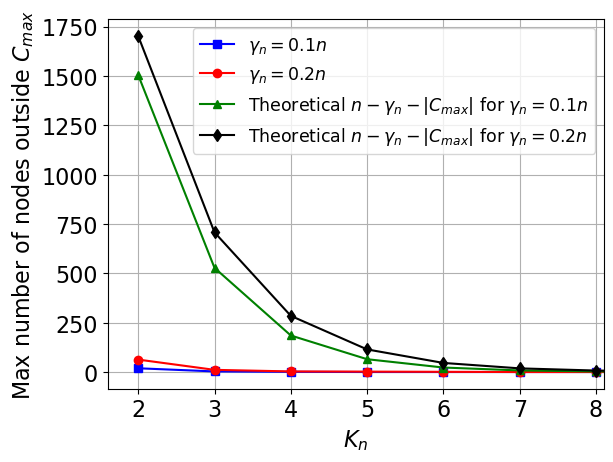}\label{fig:gc2}
\hspace{0.4mm}
\includegraphics[scale=0.45]{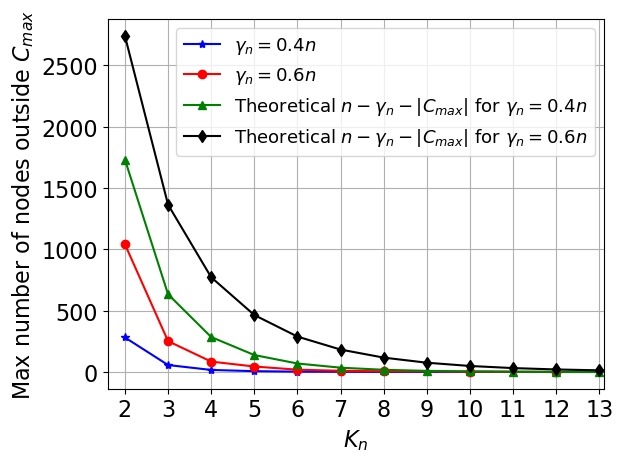}\label{fig:gc2a}  \vspace{-2mm}
\caption{\sl Maximum number of nodes outside the giant component of $\hhdn$ for $n = 5000$ and $\gamma_n = 0.1n$, $\gamma_n = 0.2n$ cases (Left); and for $n = 5000$ and $\gamma_n = 0.4n$, $\gamma_n = 0.6n$ cases (Right), obtained through 1000 experiments along with respective plot of theoretical $n - \gamma_n - |C_{max}|$.} 
\label{fig:fig3}
\end{figure}

In the first setup, we consider the case where the number of deleted nodes, $\gamma_n = \alpha n$, with $\alpha$ in $(0,1)$. We generate instantiations of the random graph $\hhdn$ with $n=5000$, varying $K_n$  in the interval $[1, 25]$ and consider several $\alpha$ values in the interval $[0.1,0.8]$. Then, we record the empirical probability of connectivity of the graph $\hhdn$ and $\lambda_n$ from 1000 independent experiments for each $(K_n,\alpha)$ pair. The results of this experiment are shown in Fig.~\ref{fig:fig1} and Fig.~\ref{fig:fig3}.
Fig.~\ref{fig:fig1} (Left) depicts the empirical probability of connectivity of $\hhdn$. The vertical lines stand for the critical threshold of connectivity obtained from Theorem \ref{theorem:thmc_1}. In each curve, $P(n,K_n,\gamma_n)$ exhibits a threshold behaviour as $K_n$ increases, and the transition from $P(n,K_n,\gamma_n)=0$ to $P(n,K_n,\gamma_n)=1$ takes place around $K_n = \frac{\log n}{1 - \alpha - \log \alpha}$, the threshold established in (\ref{eq:threshold_1}), reinforcing the usefulness of Theorem \ref{theorem:thmc_1} under practical settings.

In Fig.~\ref{fig:fig3}, the {\em maximum} number of nodes outside the giant component in 1000 experiments is plotted for each parameter pair. For comparison, we also plot the upper bound on $n - \gamma_n -|C_{max}|$ obtained from Theorem \ref{theorem:thmg_5} by taking the maximum $\gamma_n$ value that gives a threshold less than or equal to the $K_n$ value tested in the simulation. As can be seen, for any $K_n$ and $\gamma_n$ value, the experimental maximum number of nodes outside the giant component is smaller than the upper bound obtained from Theorem \ref{theorem:thmg_5}, validating the usefulness of this  result in the finite node regime.



\begin{figure}[!t]
\vspace{-3mm}
\centering
\includegraphics[scale=0.43]{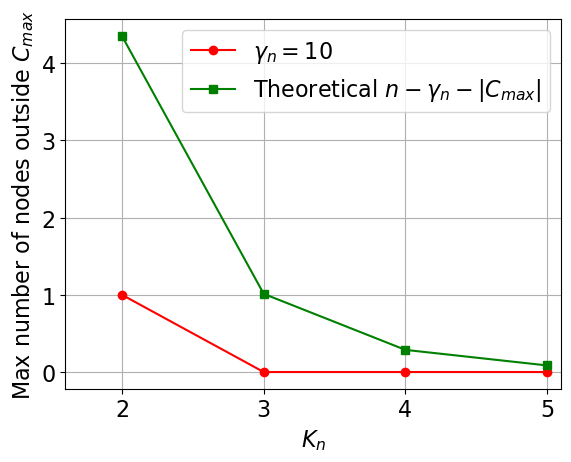}\label{fig:gc1} 
\hspace{0.4mm}
\includegraphics[scale=0.43]{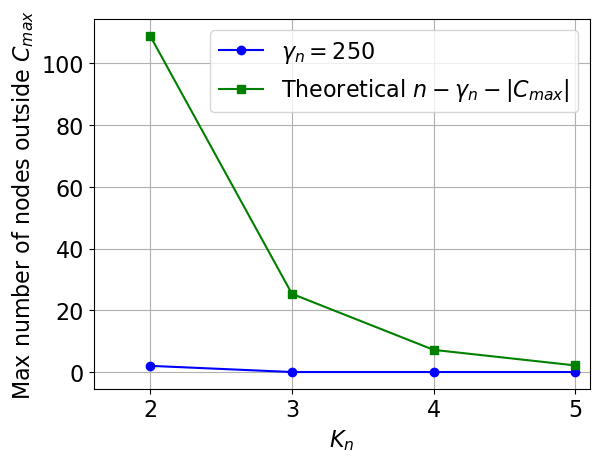}\label{fig:gc3} \vspace{-4mm}
\caption{\sl Maximum number of nodes outside the giant component of $\hhdn$ for $n = 50,000$ and $\gamma_n = 10$ cases (Left); and for for $n = 50,000$ and $\gamma_n = 250$ cases (Right), obtained through 1000 experiments along with the plot of theoretical $n - \gamma_n - |C_{max}|$. \vspace{-2mm}} 
\label{fig:fig2}
\end{figure}

The goal of the second experimental setup is to examine the case where the number of deleted nodes is $\gamma_n =  o(n)$. As before, we generate  instantiations of the random graph  $\hhdn$,
 with $n=50,000$, varying $K_n$ in  $[2, 5]$, and varying $\lambda_n$ in $[10,2000]$. For each $(K_n,\gamma_n)$ pair, the maximum number of nodes outside the giant component  in 1000 experiments is recorded; if no node is outside the giant component, then it is understood that the graph is connected.
In Fig.~\ref{fig:fig1} (Right), the maximum number of nodes outside the giant component observed in $1000$  experiments is depicted as a function of $K_n$. The plots for $\gamma_n=10$ and $\gamma_n=100$ are considered to represent  the case $\gamma_n = o(\sqrt{n})$  in Theorem \ref{theorem:thmc_2}(a). We see that there is at most one node outside the giant component for $\gamma_n=10$ and $\gamma_n=100$, even when $K_n = 2$. This shows that the asymptotic behavior given in Theorem \ref{theorem:thmc_2}(a), i.e., that random K-out graph remains connected if $o(\sqrt{n}$ nodes are deleted, already appears when $n=50,000$. The plots for $\gamma_n=1000$ and $\gamma_n=2000$ are used to check the case $\gamma_n = w(\sqrt{n})$ and $\gamma_n = o(n)$  in Theorem \ref{theorem:thmc_2}(b). The thresholds on $K_n$ for these $\gamma_n$ values, obtained using Theorem \ref{theorem:thmc_2}(b) are $r_2(1000)=6.79$ and $r_2(2000)=7.37$, rounded to two digits after decimal (the $\omega(1)$ term in Theorem \ref{theorem:thmc_2}(b) is ignored due to $n$ having a finite value in the simulations). It is clear from the plot that when $K_n \geq 4$, the graph with $\gamma_n=1000$ is connected, while   $K_n \geq 5$ suffices to ensure connectivity when $\gamma_n=2000$. Thus, selecting $K_n$ above the theoretical thresholds given in \ref{theorem:thmc_2}(b) is seen to ensure the connectivity of the graph in the finite node regime as well,  supporting the usefulness of Theorem \ref{theorem:thmc_2}(b) in practical cases.

In Fig.~\ref{fig:fig2}, the maximum number of nodes outside the giant component in 1000 experiments is plotted as a function of $K_n$.  For comparison, we also plot the upper bound on $n - |C_{max}|$ obtained from Theorem \ref{theorem:thmg_5}. In particular, for each Theorem, the maximum $\gamma_n$ value that gives a threshold less than or equal to the $K_n$ value tested in the simulation is found. Then, the lowest of these maximum $\gamma_n$ values is used as the theoretical $n - |C_{max}|$ value. As can be seen, for any $K_n$ and $\gamma_n$ value, the experimental maximum number of nodes outside the giant component is smaller than the upper bounds obtained from Theorem \ref{theorem:thmg_5}, reinforcing the usefulness of our results in the finite node regime.




\section{Proofs of Main Results}
\label{sec:proof}

In this section, we provide the proof of all  Theorems presented in Section \ref{sec:Main Results}.

\subsection{Preliminary Steps for Proving Theorems  \ref{theorem:thmc_1} - \ref{theorem:thmg_5}} \label{sec:proof_start}

 Since the preliminary steps related to the proofs of \ref{theorem:thmc_1} - \ref{theorem:thmg_5} are the same, in this Section we  present these steps.
First, recall from Section \ref{sec:model} that the metrics connectivity  and  the  size  of  the  giant  component  under node removals are defined for the graph $\hhdn$, where the set $D$ of nodes is removed from the graph $\hhn$; also recall that $R=\mathrm{V} \setminus  D$.  Let $\nodes_R$ denote the set of labels of the vertex set of $\hhdn$ and let $\mathcal{E}_n (K_n, \gamma_n; S)$ denote the event that  $S \subset \nodes_R$ is a cut in $\hhdn$ as per Definition~\ref{def:cut}. The event $\mathcal{E}_n (K_n, \gamma_n; S)$ occurs if no nodes in $S$ pick neighbors in $S\comp$, and no nodes in $S$ pick neighbors in $S\comp$. Note that nodes in $S$ or $S\comp$ can still pick neighbors in the set $\nodes_D$. Thus, we have
\begin{align}
\mathcal{E}_n (K_n, \gamma_n; S) =
\bigcap_{i \in S} \bigcap_{j \in S\comp}
\left(
\left \{ i \not \in \Gamma_{n-\gamma_n,j} \right \}
\cap 
\left \{ j \notin \Gamma_{n-\gamma_n,i} \right \}
\right). \nonumber
\end{align}

Let $\mathcal{Z}(\lambda_n;K_n, \gamma_n)$ denote the event that $\hhdn$ has no cut $S \subset \nodes_R $ with size  $\lambda_n \leq |S| \leq n-\gamma_n - \lambda_n$ where  $x:\N_0 \rightarrow  \N_0$ is a sequence such that $\lambda_n \leq (n-\gamma_n)/{2} \ \forall n$. 
In other words, $\mathcal{Z}(\lambda_n;K_n, \gamma_n)$ is the event that there are no cuts in $\hhdn$ whose size falls in the range $[\lambda_n, n-\gamma_n-\lambda_n]$. 
\begin{lemma}
\cite[Lemma 4.3]{sood2020size} For any sequence $x: \mathbb{N}_0 \rightarrow \mathbb{N}_0$ such that $\lambda_n \leq \lfloor (n-\gamma_n)/3 \rfloor$ for all $n$, we have
\begin{align}
    \mathcal{Z}(\lambda_n;K_n, \gamma_n) \Rightarrow |C_{max}(n, K_n, \gamma_n)| > n - \gamma_n - \lambda_n.
\end{align}
\label{lemma:gc}
\end{lemma}
\vspace{-5mm}
Lemma \ref{lemma:gc} states that if the event $\mathcal{Z}(\lambda_n;K_n, \gamma_n)$  holds, then the size of the largest connected component of $\hhdn$ is greater than $n - \gamma_n - \lambda_n$; i.e.,  there are less than $\lambda_n$ nodes outside of the giant component of $\hhdn$. Also note that $\hhdn$ is connected if $\mathcal{Z}(\lambda_n;K_n, \gamma_n)$ takes place with $\lambda_n=1$, since a graph is connected if no node is outside the giant component. In order to establish the Theorems \ref{theorem:thmc_1}-\ref{theorem:thmg_5}., we need to show that $\lim_{n \to \infty} \pr [\mathcal{Z}(\lambda_n;K_n, \gamma_n)\comp] = 0$ with $\lambda_n$, $K_n$ and $\gamma_n$ values as stated in each Theorem. 
From the definition of $\mathcal{Z}(\lambda_n;K_n, \gamma_n)$, we have
\begin{align}
\mathcal{Z}(\lambda_n;K_n, \gamma_n) & = \bigcap_{S \in \mathcal{P}_n: ~\lambda_n\leq  |S| \leq \lfloor \frac{n-\gamma_n}{2} \rfloor}  \left(\mathcal{E}_n({K}_n,{\gamma_n}; S)\right)\comp, \nonumber
\end{align}
where $\mathcal{P}_n$ is the collection of all non-empty  subsets of $\nodes_R$. Complementing both sides and using the union bound, we get
\begin{align}
\pr\left[\left(\mathcal{Z}(\lambda_n;K_n, \gamma_n)\right)\comp\right] &\leq \hspace{-3mm}  \sum_{ S \in \mathcal{P}_n: \lambda_n \leq |S| \leq \lfloor \frac{n-\gamma}{2} \rfloor } \pr[ \mathcal{E}_n ({K}_n,{\gamma_n}; S) ] \nonumber \\
&=\hspace{-1mm} \sum_{r=\lambda_n}^{ \left\lfloor \frac{n-\gamma}{2} \right\rfloor } \hspace{-1mm}
 \sum_{S \in \mathcal{P}_{n,r} } \pr[\mathcal{E}_n ({K}_n,{\gamma_n}; S)] \label{eq:BasicIdea+UnionBound},
\end{align}
where  $\mathcal{P}_{n,r} $ denotes the collection of all subsets of $\nodes_R$ with exactly $r$ elements.
For each $r=1, \ldots , \left\lfloor (n-\gamma_n)/2\right\rfloor$, we can simplify the notation by denoting $\mathcal{E}_{n,r} ({K}_n,{\gamma_n})=\mathcal{E}_n ({K}_n,{\gamma_n} ; \{ 1, \ldots , r \} )$. From the exchangeability of the node labels and associated random variables, we have
\[
\pr[ \mathcal{E}_n({K}_n,{\gamma_n} ; S) ] = \pr[ \mathcal{E}_{n,r}({K}_n,{\gamma_n}) ], \quad S \in
\mathcal{P}_{n,r}.
\]
$|\mathcal{P}_{n,r} | = {n-\gamma_n \choose r}$, since there are ${n-\gamma_n \choose r}$ subsets of $\nodes_R$ with r elements. Thus, we have
\begin{equation*}
\sum_{S \in \mathcal{P}_{n,r} } \pr[\mathcal{E}_n ({K}_n,{\gamma_n} ; S) ] 
= {n-\gamma_n\choose r} ~ \pr[\mathcal{E}_{n,r} ({K}_n,{\gamma_n})]. 
\label{eq:ForEachr}
\end{equation*}
Substituting this into (\ref{eq:BasicIdea+UnionBound}), we obtain 
\begin{align}
\hspace{-.4mm} \pr\left[\left(\mathcal{Z}(\lambda_n;K_n,\gamma_n)\right)\comp\right] \leq \hspace{-1mm} \sum_{r=\lambda_n}^{ \left\lfloor \frac{\hspace{-.5mm} n-\gamma}{2} \right\rfloor } \hspace{-1mm} 
{n-\gamma_n \hspace{-.4mm} \choose r } \hspace{-.4mm}  \pr[ \mathcal{E}_{n,r}({K}_n,{\gamma_n})] \hspace{-1mm} 
\label{eq:Z_bound}
\end{align}

Remember that $\mathcal{E}_{n,r}({K}_n,{\gamma_n})$ is the event that the $n-\gamma_n-r$ nodes in $S$ and $r$ nodes in $S\comp$ do not pick each other, but they can pick nodes from the set $\nodes_D$. Thus, we have:
 \begin{align}
\pr [\mathcal{E}_{n,r}({K}_n,{\gamma_n})] & =   \left( \dfrac{{\gamma_n+r-1 \choose K_n}}{{n-1 \choose K_n}} \right)^{r} \left( \dfrac{{n-r-1 \choose K_n}}{{n-1 \choose K_n}} \right)^{n-\gamma_n-r} \leq \left(\dfrac{\gamma_n+r}{n}\right)^{rK_n} 
\left(\dfrac{n-r}{n}\right)^{K_n(n-\gamma_n-r)} 
\end{align}
Abbreviating $\pr\left[\mathcal{Z}(1;K_n, \gamma_n)\comp\right]$ as $P_Z$, we get from (\ref{eq:Z_bound}) that
\begin{align}
\hspace{-1mm} P_Z \leq \hspace{-1mm}  \sum_{r=\lambda_n}^{ \left\lfloor \frac{n-\gamma_n}{2} \right\rfloor } \hspace{-1mm} {\hspace{-.5mm}n-\gamma_n\hspace{-.5mm}\choose r} \hspace{-1mm} \left(\hspace{-.5mm}\dfrac{\hspace{-.5mm}\gamma_n+r\hspace{-.5mm}}{n}\right)^{\hspace{-1mm} r K_n} 
\hspace{-1mm} \left(\hspace{-.5mm}\dfrac{n-r}{n}\hspace{-.5mm} \right)^{\hspace{-1mm} K_n(n-\gamma_n-r)\hspace{-.5mm}} 
\label{eq:gc_pz0}
\end{align}
Using the upper bound on binomials \eqref{eq:binomial_property} again, we have
\begin{align}
P_Z \leq \sum_{r=\lambda_n}^{ \left\lfloor \frac{n-\gamma_n}{2} \right\rfloor } \left( \frac{n-\gamma_n}{r}\right)^r \left( \frac{n-\gamma_n}{n-\gamma_n -r}\right)^{n-\gamma_n-r}  \left(\dfrac{\gamma_n +r}{n}\right)^{rK_n} 
\left(\dfrac{n-r}{n}\right)^{K_n(n-\gamma_n -r)} 
\label{eq:gc_pz}
\end{align}

In order to establish the Theorems, we need to show that (\ref{eq:gc_pz}) goes to zero in the limit of large n for $\lambda_n$, $\gamma_n$ and $K_n$ values as specified in each Theorem. 

Since they will be referred to frequently throughout the proofs, we also include here the following standard bounds.
\begin{align}
    1\pm x \leq e^{\pm x} \label{eq:expon_upper}
\end{align}
\vspace{-2mm}
\begin{align}
{n \choose m} \leq \left(\frac{n}{m}\right)^m \left(\frac{n}{n-m}\right)^{n-m}, \quad \forall m=1,\ldots,n
\label{eq:binomial_property} 
\end{align}

\subsection{A Proof of Theorem \ref{theorem:thmc_1}}

Recall that in Theorem \ref{theorem:thmc_1}, we have $\gamma_n = \alpha n$ with $0<\alpha<1$ and that we need  $\lambda_n=1$ for connectivity. Using  \eqref{eq:expon_upper}
in  (\ref{eq:gc_pz}), we have
\begin{align}
P_Z \leq \sum_{r=1}^{ \left\lfloor \frac{n-\alpha n}{2} \right\rfloor } \left(\dfrac{n -\alpha n}{r}\right)^r e^r \left(\alpha + \dfrac{r}{n}\right)^{rK_n} e^{\frac{-r K_n (n-\alpha n -r)}{n}} \nonumber
\end{align}
We will show that the right side of the above expression goes to zero as $n$ goes to infinity. Let
 $$A_{n,r,\alpha}: = \left(\dfrac{n -\alpha n}{r}\right)^r e^r \left(\alpha + \dfrac{r}{n}\right)^{rK_n}e^{\frac{-rK_n(n-\alpha n -r)}{n}}.$$
We write
\begin{align*}
P_Z \leq \sum_{r=1}^{ \left\lfloor n/\log n \right\rfloor } A_{n,r,\alpha} + \sum_{r=\left\lfloor n/\log n \right\rfloor}^{ \left\lfloor \frac{n-\alpha n}{2} \right\rfloor } A_{n,r,\alpha} := S_1 + S_2,
\end{align*}
and show that both $S_1$ and $S_2$ go to zero as $n \to \infty$.
We start with the first summation $S_1$.
\begin{align}
S_1  & \! \begin{multlined}[t]
 \leq \sum_{r=1}^{ \left\lfloor n/\log n  \right\rfloor } \left((1 -\alpha )en\cdot e^{  K_n \log(\alpha+ \frac{1}{\log n})  -K_n (1-\alpha -\frac{1}{\log n}) } \right)^r \nonumber
 \end{multlined}
 \end{align}
Next, assume as in the statement of Theorem \ref{theorem:thmc_1} that 
\begin{align}
    K_n = \frac{c_n \log n}{1 - \alpha - \log \alpha}, \quad n=1,2,\ldots
\label{eq:proof1_k}
\end{align}
 for some sequence $c: \mathbb{N}_0 \to \mathbb{R}_+$ such that $\lim_{n \to \infty}c_n = c$ with $c>1$. 
Also define 
\begin{align*}
a_n &:=  (1 -\alpha )en\cdot e^{  K_n\log(\alpha+ \frac{1}{\log n})  -K_n(1-\alpha -\frac{1}{\log n}) }  \\
& = (1 -\alpha )en\cdot e^{  -\frac{c_n \log n}{1 - \alpha - \log \alpha}\left(1-\alpha -\frac{1}{\log n} - \log(\alpha+ \frac{1}{\log n}) \right) } 
\\
& = (1 -\alpha )e n^{1-c_n} \cdot e^{\frac{c_n}{1- \alpha - \log \alpha}\left(1- \log n \cdot \log(1+ \frac{1}{\alpha \log n}) \right)}
\\
& = O(1) n^{1-c_n}
\end{align*}
where we substituted $K_n$ via (\ref{eq:proof1_k}) and used the fact that $ \log n \cdot \log(1+ \frac{1}{\alpha \log n}) =  \frac{1}{\alpha} +o(1)$. Taking the limit as $n \to \infty$ and recalling that 
$\lim _{n \to \infty} c_n = c >1$, we see that $\lim _{n \to \infty} a_n = 0$. Hence, for large $n$, we  have
\begin{align}
S_1 \leq \sum_{r=1}^{ \left\lfloor n/\log n  \right\rfloor } \left( a_n \right)^r \leq \sum_{r=1}^{ \infty} \left( a_n \right)^r = \frac{a_n}{1-a_n}
\end{align}
where the geometric sum converges by virtue of $\lim _{n \to \infty} a_n = 0$. Using this, it is clear that 
$\lim _{n \to \infty}S_1 = 0$.

Now, consider the second summation $S_2$.
\begin{align}
S_2 & \leq \! \begin{multlined}[t]  \sum_{r= \lfloor n/\log n\rfloor }^{ \left\lfloor (n-\alpha n)/2 \right\rfloor } \left(\dfrac{(n -\alpha n )e}{n/\log n}\right)^r 
\left(\dfrac{\alpha n +\frac{n-\alpha n}{2} }{n}\right)^{r K_n}  e^{\frac{-rK_n}{n}(n-\alpha n - \frac{n-\alpha n}{2})} 
\end{multlined} \\
& \! \begin{multlined}
 \leq \sum_{r=  \lfloor n/\log n\rfloor }^{ \left  \lfloor (n-\alpha n)/2 \right\rfloor } \left((1 - \alpha )e\log n \cdot e^{ K_n \log(\frac{1+\alpha}{2}) -K_n\frac{1- \alpha}{2} } \right)^r 
  \end{multlined} 
\end{align}
Next, we define 
\begin{align}
b_n &:= (1 - \alpha )e\log n \cdot e^{ -K_n \left(  \frac{1-\alpha}{2}  - \log(\frac{1+\alpha}{2}) \right)} \\
& = (1 - \alpha )e\log n \cdot e^{ -\frac{c_n \log n}{1 - \alpha - \log \alpha} \left(  \frac{1-\alpha}{2}  - \log(\frac{1+\alpha}{2}) \right)}
\end{align}
where we substituted for $K_n$ via (\ref{eq:proof1_k}). Taking the limit as ${n \to \infty}$ we see that
$
\lim _{n \to \infty} b_n = 0
$
upon noting that $ \frac{1-\alpha}{2}  - \log(\frac{1+\alpha}{2}) >0$ and $\lim _{n \to \infty} c_n =c >1$.
 With  arguments similar to those used in the case of $S_1$, we can show that when $n$ is large, $S_2 \leq b_n/(1-b_n)$, leading to $S_2$ converging to zero as $n$ gets large.
With $P_Z \leq S_1 + S_2$, and both $S_1$ and $S_2$ converging to zero when $n$ is large, we establish the fact that $P_Z$ converges to zero as $n$ goes to infinity. This result also yields the desired conclusion $\lim _{n \to \infty} P(n,K_n,\gamma_n) =1$ in Theorem \ref{theorem:thmc_1} since $P_Z = 1-P(n,K_n,\gamma_n)$.

\subsection{A Proof of Theorem \ref{theorem:thmc_2}}
We will first start with part (a) of Theorem \ref{theorem:thmc_2}. \\
\textbf{Part a)} Recall that in part (a), $\gamma_n = o(\sqrt{n})$ and  we need $\lambda_n = 1$ for connectivity. Using this and \eqref{eq:expon_upper} in \eqref{eq:gc_pz}, we get
\begin{align}
P_Z & \leq \! \begin{multlined}[t] \sum_{r=1}^{ \left\lfloor \frac{n-\gamma_n}{2} \right\rfloor }  \left(\dfrac{n -\gamma_n}{r}\right)^r \left( \frac{n-\gamma_n}{n-\gamma_n -r}\right)^{n-\gamma_n-r}   \left( \dfrac{\gamma_n + r}{n}\right)^{r K_n} \left(\frac{n-r}{n}\right)^{K_n(n-\gamma_n -r)}   \nonumber
\end{multlined} \nonumber \\
& \leq \! \begin{multlined}[t] \sum_{r=1}^{ \left\lfloor \frac{n-\gamma_n}{2} \right\rfloor }  \left(\dfrac{n -\gamma_n}{r}\right)^r \left( 1 + \frac{r \gamma_n}{n(n-\gamma_n -r)}\right)^{n-\gamma_n-r}  \left( \dfrac{\gamma_n + r}{n}\right)^{r K_n} \left(\frac{n-r}{n}\right)^{(K_n-1)(n-\gamma_n -r)}   \nonumber
\end{multlined} \nonumber \\
& \leq \! \begin{multlined}[t] \sum_{r=1}^{ \left\lfloor \frac{n-\gamma_n}{2} \right\rfloor }  \left( 1+\dfrac{\gamma_n }{r}\right)^{r}    \left( \dfrac{\gamma_n + r}{n}\right)^{r (K_n-1)}  e^{\frac{-r(K_n-1)(n-\gamma_n -r)}{n}}   \nonumber 
\end{multlined} \nonumber 
\end{align}
We will show that the right side of the above expression goes to zero as $n$ goes to infinity. Let
 $$A_{n,r,\gamma_n}: = \left( 1+\dfrac{\gamma_n }{r}\right)^{r}   \left( \dfrac{\gamma_n + r}{n}\right)^{r (K_n-1)} e^{\frac{-r(K_n-1)(n-\gamma_n -r)}{n}} $$
We write
\begin{align*}
P_Z \leq \sum_{r=1}^{ \left\lfloor \sqrt{n} \right\rfloor } A_{n,r,\gamma_n} + \sum_{r=\left\lceil \sqrt{n} \right\rceil}^{ \left\lfloor \frac{n-\gamma_n}{2} \right\rfloor } A_{n,r,\gamma_n} := S_1 + S_2,
\end{align*}
and show that both $S_1$ and $S_2$ go to zero as $n \to \infty$.
We start with the first summation $S_1$.
\begin{align}
S_1 & \leq \sum_{r=1}^{ \left\lfloor \sqrt{n} \right\rfloor }  \left( 1+\frac{\gamma_n }{r}\right)^{r}    \left( \frac{\gamma_n + r}{n}\right)^{r (K_n-1)}  e^{\frac{-r(K_n-1)(n-\gamma_n -r)}{n}} \nonumber \\
& \leq \! \begin{multlined}[t] \sum_{r=1}^{ \left\lfloor \sqrt{n} \right\rfloor }  \left(e^{\log \left( 1+\gamma_n \right)  + (K_n-1) \left[  \log \left( \frac{\gamma_n + \sqrt{n}}{n}\right) - \frac{n-\gamma_n -\sqrt{n}}{n}\right] }\right)^r \nonumber
\end{multlined} \nonumber
\end{align}
Next, assume as in the statement of Theorem \ref{theorem:thmc_2}(a) that \\ $K_n\geq 2, \  \forall n$. Also define
\begin{align*}
a_n &:= e^{\log \left( 1+\gamma_n \right)  + (K_n-1) \left[  \log \left( \frac{\gamma_n + \sqrt{n}}{n}\right) - \frac{n-\gamma_n -\sqrt{n}}{n}\right] }
\\
& \leq e^{\log \left( 1+\gamma_n \right)  + \log \left( 1+\frac{\gamma_n}{\sqrt{n}} \right) -  \log \left( \sqrt{n}\right)} e^{- \frac{n-\gamma_n -\sqrt{n}}{n} }
\\
& = O(1) e^{\log \left( 1+\gamma_n \right) -  \log \left( \sqrt{n}\right)}
\end{align*}
Taking the limit as $n \to \infty$ and recalling that $\gamma_n = o(\sqrt{n})$, we see that
$\lim _{n \to \infty} a_n = 0$. Hence, for large $n$, we  have
\begin{align}
S_1 \leq \sum_{r=1}^{ \left\lfloor \sqrt{n}  \right\rfloor } \left( a_n \right)^r \leq \sum_{r=1}^{ \infty} \left( a_n \right)^r = \frac{a_n}{1-a_n}
\end{align}
where the geometric sum converges by virtue of $\lim _{n \to \infty} a_n = 0$. Using this once again, it is clear from the last expression that 
$\lim _{n \to \infty}S_1 = 0$.

Now, consider the second summation $S_2$.
\begin{align}
S_2 & \leq \! \begin{multlined}[t] \sum_{r=\left\lceil \sqrt{n} \right\rceil}^{ \left\lfloor \frac{n-\gamma_n}{2} \right\rfloor }  \left(e^{\frac{r\gamma_n}{\sqrt{n}}  + (K_n-1) \left[  \log \left( \frac{n + \gamma_n }{2n}\right) - \frac{n-\gamma_n }{2n}\right] }\right)^r \nonumber
\end{multlined} \nonumber
\end{align}

Again assume as in the statement of Theorem \ref{theorem:thmc_2}(a) that $K_n\geq 2$. Next, we define=
\begin{align*}
b_n &:= e^{\frac{r\gamma_n}{\sqrt{n}}  + (K_n-1) \left[  \log \left( \frac{n + \gamma_n }{2n}\right) - \frac{n-\gamma_n }{2n}\right] }
\\
& \leq e^{\frac{r\gamma_n}{\sqrt{n}}  +   \log \left( \frac{1 }{2}\right) +   \log \left( \frac{n }{n+\gamma_n}\right) - \frac{1}{2} + \frac{\gamma_n }{2n} }
\\
& = O(1) e^{-  \log \left( \sqrt{n}\right)}
\end{align*}
Taking the limit as $n \to \infty$ and recalling that $\gamma_n = o(\sqrt{n})$, we see that
$\lim _{n \to \infty} b_n = 0$. Hence, for large $n$, we  have
\begin{align}
S_2 \leq \sum_{r=\left\lceil \sqrt{n} \right\rceil}^{ \left\lfloor \frac{n-\gamma_n}{2} \right\rfloor }  \left( b_n \right)^r \leq \sum_{r=\left\lceil \sqrt{n} \right\rceil}^{ \infty} \left( b_n \right)^r = \frac{(b_n)^{\sqrt{n}}}{1-b_n}
\end{align}
where the geometric sum converges by virtue of $\lim _{n \to \infty} b_n = 0$. Using this once again, it is clear from the last expression that 
$\lim _{n \to \infty}S_2 = 0$. With $P_Z \leq S_1 + S_2$, and both $S_1$ and $S_2$ converging to zero when $n$ is large, we establish the fact that $P_Z$ converges to zero as $n$ goes to infinity. This result also yields the desired conclusion $\lim _{n \to \infty} P(n,K_n,\gamma_n) =1$ in Theorem \ref{theorem:thmc_2}(a) since $P_Z = 1-P(n,K_n,\gamma_n)$. 

\vspace{4mm}
\textbf{Part b)} We now continue with the proof of Theorem \ref{theorem:thmc_2}(b). Recall that we had $\gamma_n = \Omega(\sqrt{n})$ and $\gamma_n = o(n)$. Using this and \eqref{eq:expon_upper}
in (\ref{eq:gc_pz}), we get
\begin{align}
P_Z & \leq \! \begin{multlined}[t] \sum_{r=1}^{ \left\lfloor \frac{n-\gamma_n}{2} \right\rfloor } \left(\dfrac{n -\gamma_n}{r}\right)^r \left( \frac{n-\gamma_n}{n-\gamma_n -r}\right)^{n-\gamma_n-r}  \left( \dfrac{\gamma_n + r}{n}\right)^{rK_n} e^{\frac{-rK_n(n-\gamma_n -r)}{n}} \nonumber \end{multlined} \nonumber \\
& \leq \! \begin{multlined}[t] \sum_{r=1}^{ \left\lfloor \frac{n-\gamma_n}{2} \right\rfloor } e^{(1-\gamma_n/n)r} \left( \dfrac{\gamma_n + r}{r}\right)^{r} \left( \dfrac{\gamma_n + r}{n}\right)^{r(K_n-1)} e^{\frac{-rK_n(n-\gamma_n -r)}{n}}
\end{multlined} \nonumber \\
& \leq \! \begin{multlined}[t] \sum_{r=1}^{ \left\lfloor \frac{n-\gamma_n}{2} \right\rfloor } \exp \left( r \left[ (K_n-1) \left( \log \left( \frac{\gamma_n + r}{n}\right) + \frac{r}{n}   + \frac{\gamma_n}{n} -1  \right)   + \log \left(1 + \gamma_n \right) 
+ \frac{n-\gamma_n}{2n} \right] \right) 
\end{multlined} \label{eq:proof3_2}
\end{align}

Next, assume as in the statement of Theorem \ref{theorem:thmc_2}(b) that
\begin{align}
    K_n = \frac{\log(\gamma_n+1)}{\log2 + 1/2} + w(1), \quad n=1,2,\ldots
\label{eq:proof3_2a}
\end{align}
Since $K_n-1 > 0, \quad \forall n=1,2,\ldots$, and noting that $r\leq \left\lfloor \frac{n-\gamma_n}{2} \right\rfloor$ in (\ref{eq:proof3_2}), we have 
\begin{align}
(K_n-1) \left( \log \left( \frac{\gamma_n + r}{n}\right) + \frac{r}{n} + \frac{\gamma_n}{n} -1  \right) \leq
(K_n-1) \left( \log \left( \frac{\gamma_n + \frac{n-\gamma_n}{2}}{n}\right) + \frac{\frac{n-\gamma_n}{2}}{n} + \frac{\gamma_n}{n} -1  \right) 
\end{align}
Using this, we get
\begin{align}
P_Z & \leq \! \begin{multlined}[t] \sum_{r=1}^{ \left\lfloor \frac{n-\gamma_n}{2} \right\rfloor } \exp \left( r \left[ (K_n-1) \left( \log \left( \frac{n + \gamma_n }{2n}\right)    - \frac{n-\gamma_n}{2n}   \right)   + \log \left(1 + \gamma_n \right) 
+ \frac{n-\gamma_n}{2n} \right] \right)
\end{multlined}
\end{align}
Next, define 
\begin{align}
    a_n:=e^{(K_n-1)\left(\log \left( \frac{n + \gamma_n}{2n}\right) - \frac{n-\gamma_n}{2n}   \right) + \log \left(1 + \gamma_n \right) 
+ \frac{n-\gamma_n}{2n}}
\end{align}
Recall that $\gamma_n=o(n)$, so we have $\lim _{n \to \infty} \gamma_n/n = 0$. Using this, and substituting $K_n$ via (\ref{eq:proof3_2a}), we get
\begin{align}
    \lim _{n \to \infty} a_n &= \lim _{n \to \infty} \left[ e^{\left(\frac{\log(\gamma_n+1)}{\log 2 +1/2} + w(1)\right)\cdot \left(-log 2 - \frac{1}{2}   \right) + \log \left(1 + \gamma_n \right) 
+ \frac{1}{2}}\right] \nonumber \\
&= \lim _{n \to \infty} \left[ e^{ -w(1)\cdot \left(log 2 + 1/2   \right) - \log \left(1 + \gamma_n \right)+ \log \left(1 + \gamma_n \right) + \frac{1}{2}}\right] \nonumber \\
&= \lim _{n \to \infty} \left[ o(1) e^{ -w(1)\cdot \left(log 2 + 1/2  \right)} \right] = 0
\end{align}
Hence, for large $n$, we  have
\begin{align}
P_Z \leq \sum_{r=1}^{ \left\lfloor \frac{n-\gamma_n}{2} \right\rfloor } \left( a_n \right)^r \leq \sum_{r=1}^{ \infty} \left( a_n \right)^r = \frac{a_n}{1-a_n}
\end{align}
where the geometric sum converges by virtue of $\lim _{n \to \infty} a_n = 0$. Using this, it is clear from the last expression that 
$\lim _{n \to \infty}P_Z = 0$. This result also yields the desired conclusion $\lim _{n \to \infty} P(n,K_n,\gamma_n) =1$ in Theorem \ref{theorem:thmc_2}(b) since $P_Z = 1-P(n,K_n,\gamma_n)$. This result, combined with the proof of part a, concludes the proof of Theorem \ref{theorem:thmc_2}.

\subsection{A Proof of Theorem \ref{theorem:thmg_3}}

Recall that in Theorem \ref{theorem:thmg_3}, we have $\gamma_n = o(n)$ and $\lambda_n = \Omega(\sqrt{n})$. Using \eqref{eq:expon_upper} in (\ref{eq:gc_pz}),  we have
\begin{align}
P_Z & \leq \! \begin{multlined}[t] \sum_{r=\lambda_n}^{ \left\lfloor \frac{n-\gamma_n}{2} \right\rfloor }  \left(\dfrac{n -\gamma_n}{r}\right)^r \left( \frac{n-\gamma_n}{n-\gamma_n -r}\right)^{n-\gamma_n-r}   \left( \dfrac{\gamma_n + r}{n}\right)^{r K_n} \left(\frac{n-r}{n}\right)^{K_n(n-\gamma_n -r)}   \nonumber
\end{multlined} \nonumber \\
& \leq \! \begin{multlined}[t] \sum_{r=\lambda_n}^{ \left\lfloor \frac{n-\gamma_n}{2} \right\rfloor }  \left(\dfrac{n -\gamma_n}{r}\right)^r \left( 1 + \frac{r \gamma_n}{n(n-\gamma_n -r)}\right)^{n-\gamma_n-r}  \left( \dfrac{\gamma_n + r}{n}\right)^{r K_n} \left(\frac{n-r}{n}\right)^{(K_n-1)(n-\gamma_n -r)}   \nonumber
\end{multlined} \nonumber \\
& \leq \! \begin{multlined}[t] \sum_{r=\lambda_n}^{ \left\lfloor \frac{n-\gamma_n}{2} \right\rfloor }  \left(\dfrac{n -\gamma_n}{r}\right)^r e^{\frac{r\gamma_n}{n}}\left( \dfrac{\gamma_n + r}{n}\right)^{r}   \left( \dfrac{\gamma_n + r}{n}\right)^{r (K_n-1)} e^{\frac{-r(K_n-1)(n-\gamma_n -r)}{n}}   \nonumber
\end{multlined} \nonumber \\
& \leq \! \begin{multlined}[t] \sum_{r=\lambda_n}^{ \left\lfloor \frac{n-\gamma_n}{2} \right\rfloor }  \left( 1+\dfrac{\gamma_n }{r}\right)^{r}  \left( \dfrac{\gamma_n + r}{n}\right)^{r (K_n-1)}  e^{\frac{-r(K_n-1)(n-\gamma_n -r)}{n}}   \nonumber 
\end{multlined} \nonumber 
\end{align}
Next, assume as in the statement of Theorem \ref{theorem:thmg_3} that
\begin{align}
    K_n > 1 + \frac{\log(1+\gamma_n / \lambda_n)}{\log 2 + 1/2}
    \label{eq:proof_3_3}
\end{align}
Since $K_n > 1$, we have
\begin{align}
P_Z & \leq \! \begin{multlined}[t] \sum_{r=\lambda_n}^{ \left\lfloor \frac{n-\gamma_n}{2} \right\rfloor }  \left( 1+\dfrac{\gamma_n }{\lambda_n}\right)^{r}    \left( \dfrac{n + \gamma_n }{2n}\right)^{r (K_n-1)}  e^{\frac{-r(K_n-1)(n-\gamma_n )}{2n}}   \nonumber
\end{multlined} \nonumber 
\end{align}
We will show that the right side of the above expression goes to zero as $n$ goes to infinity. Let
 $$a_n: = e^{\log\left(1+\frac{\gamma_n }{\lambda_n}\right) + (K_n-1)\left[ \log\left(\frac{n+\gamma_n}{2n}\right) - \frac{n-\gamma_n}{2n}\right]} $$
Recall that $\gamma_n=o(n)$, so we have $\lim_{n\to \infty}\gamma_n/n=0$. Using this, and substituting for $K_n$ via (\ref{eq:proof_3_3}), we get
\begin{align}
    a_n & < e^{\log\left(1+\frac{\gamma_n }{\lambda_n}\right) + \left( \frac{\log(1+\gamma_n / \lambda_n)}{- \log(1/2) - 1/2}\right)\left[ \log\left(\frac{n+\gamma_n}{2n}\right) - \frac{n-\gamma_n}{2n}\right]}  
\end{align}
Taking the limit $n\to \infty$, we have
\begin{align}
    \lim_{n\to \infty} a_n & <   \lim_{n\to \infty} e^{\log\left(1+\frac{\gamma_n }{\lambda_n}\right) -  \log(1+\gamma_n / \lambda_n)} = e^0 = 1 
\end{align}
Hence, for large $n$, we  have
\begin{align}
P_Z \leq \sum_{r=\lambda_n}^{ \left\lfloor \frac{n-\gamma_n}{2} \right\rfloor } \left( a_n \right)^r \leq \sum_{r=\lambda_n}^{ \infty} \left( a_n \right)^r = \frac{(a_n)^{\lambda_n}}{1-a_n}
\end{align}
where the geometric sum converges by virtue of $\lim _{n \to \infty} a_n < 1$ and $\lim _{n \to \infty} \lambda_n = w(1)$. Using this, it is clear from the last expression that 
$\lim _{n \to \infty}P_Z = 0$. This result also yields the desired conclusion $\lim _{n \to \infty} P_G(n,K_n,\gamma_n,\lambda_n) =1$ in Theorem \ref{theorem:thmg_3} since $P_Z = 1-P_G(n,K_n,\gamma_n,\lambda_n)$. This concludes the proof of Theorem \ref{theorem:thmg_3}.

\subsection{A Proof of Theorem \ref{theorem:thmg_5}}

Recall that in Theorem \ref{theorem:thmg_5}, we have $\gamma_n = \alpha n$ with $\alpha$ in $(0,1)$, and $\lambda_n < \frac{(1-\alpha)n}{2}$. Using $\gamma_n = \alpha n$  in (\ref{eq:gc_pz}), we get
\begin{align}
P_Z & \leq \! \begin{multlined}[t] \sum_{r=\lambda_n}^{ \left\lfloor \frac{n-\alpha n}{2} \right\rfloor }  \left(\dfrac{n -\alpha n}{r}\right)^r \left( \frac{n-\alpha n}{n-\alpha n -r}\right)^{n-\alpha n-r}   \left( \dfrac{\alpha n + r}{n}\right)^{r K_n} \left(\frac{n-r}{n}\right)^{K_n(n-\alpha n -r)}   \nonumber
\end{multlined} \nonumber \\
& \leq \! \begin{multlined}[t] \sum_{r=\lambda_n}^{ \left\lfloor \frac{n-\alpha n}{2} \right\rfloor }  \left(1-\alpha \right)^r e^{\alpha r}\left( 1+\dfrac{\alpha n }{r}\right)^{r} \left( \dfrac{\alpha n + r}{n}\right)^{r (K_n-1)} e^{-r(K_n-1)\frac{(n-\alpha n -r)}{n}}    \\
\end{multlined} 
\label{eq:proof3_5_2}
\end{align}
Next, assume as in the statement of Theorem \ref{theorem:thmg_5} that 
\begin{align}
    K_n > 1 + \frac{ \log(1+\frac{\alpha n}{\lambda_n}) + \alpha + \log(1-\alpha) }{\frac{1-\alpha}{2} +  \log2 - \log(1+\alpha)   }, \quad n=1,2,\ldots \nonumber
\label{eq:proof3_5a}
\end{align}
Since $K_n > 1$, we have
\begin{align}
P_Z  
& \leq \! \begin{multlined}[t] \sum_{r=\lambda_n}^{ \left\lfloor \frac{n-\alpha n}{2} \right\rfloor }  \left(1-\alpha \right)^r e^{\alpha r}\left( 1+\dfrac{\alpha n }{\lambda_n }\right)^{r}  \left( \dfrac{1 + \alpha}{2}\right)^{r (K_n-1)}   e^{-r(K_n-1)\left( \frac{1-\alpha}{2}\right)} 
\end{multlined} \nonumber
\end{align}
Also define 
\vspace{1mm}
\begin{align}
a_n &:=  e^{\alpha +  \log(1-\alpha) + \log(1+\frac{\alpha n}{\lambda_n}) + (K_n-1)\left[  \log \left(\frac{1+\alpha}{2}\right) - \left( \frac{1-\alpha}{2}\right) \right] } \nonumber \\
&<  e^{\alpha +  \log(1-\alpha) + \log(1+\frac{\alpha n}{\lambda_n}) - \left( \alpha +  \log(1-\alpha) + \log(1+\frac{\alpha n}{\lambda_n}) \right) } = 1 \nonumber
\end{align}
where we substituted $K_n$ via (\ref{eq:proof3_5a}). Taking the limit as $n \to \infty$, we see that $\lim_{n \to \infty} a_n < 1$. Hence, for large n, we have
\begin{align}
P_Z \leq \sum_{r=\lambda_n}^{ \left\lfloor \frac{n-\alpha n}{2} \right\rfloor } \left( a_n \right)^r \leq \sum_{r=\lambda_n}^{ \infty} \left( a_n \right)^r = \frac{(a_n)^{\lambda_n}}{1-a_n}
\end{align}
where the geometric sum converges by virtue of $\lim _{n \to \infty} a_n < 1$ and $\lim _{n \to \infty} \lambda_n = w(1)$. Using this, it is clear from the last expression that 
$\lim _{n \to \infty}P_Z = 0$. This result also yields the desired conclusion $\lim _{n \to \infty} P_G(n,K_n,\gamma_n,\lambda_n) =1$ in Theorem \ref{theorem:thmg_5} since $P_Z = 1-P_G(n,K_n,\gamma_n,\lambda_n)$. This concludes the proof of Theorem \ref{theorem:thmg_5}.

\subsection{Preliminaries Needed in the  Proof of Theorem \ref{theorem:thmr_1} }

We start with a few definitions and properties that will be useful throughout the rest of the proof. First, let $B(a,b)$ denote the beta function, $B_x(a,b)$ denote the incomplete beta function, and $I_x(a,b)$ denote the regularized incomplete beta function, where $a$ and $b$ are non-negative integers. These functions are defined as follows \cite{NIST:DLMF}:
\begin{align}
    B(a,b)&= \int_{0}^{1} t^{a-1}(1-t)^{b-1} dt = \frac{(a-1)!(b-1)!}{(a+b-1)!} \nonumber \\
    B_x(a,b)&= \int_{0}^{x} t^{a-1}(1-t)^{b-1} dt, \quad 0\leq x \leq 1 \nonumber \\
    I_x(a,b)&=\frac{B_x(a,b)}{B(a,b)}, \quad 0\leq x \leq 1
\end{align}

Using these definitions, it can easily be shown that 
 \begin{align}
     I_{1/2}(r,r) = 1/2, \quad r > 0
 \end{align}
Proof: $
    B(r,r) =  \int_{0}^{1} t^{r-1}(1-t)^{r-1}dt  = 2 \int_{0}^{\frac{1}{2}} t^{r-1}(1-t)^{r-1}dt 
     = 2 B_{1/2}(r,r)  $
  where we divided the integral into two parts since the function $(t-t^2)^{r-1}$ is symmetric around 1/2.
 Using the fact that $B_{1/2}(r,r) = I_{1/2}(r,r) B(r,r)$, we can conclude that  $I_{1/2}(r,r) = 1/2$.

The cumulative distribution function $F(a;n,p)$ of a Binomial random variable $X \sim B(n,p)$ can be expressed using the regularized incomplete beta function as:
\begin{align}
    F(a;n,p) &= \pr[X \leq a] = I_{1-p}(n-a,a+1) \nonumber \\
    &=(n-a){n \choose a} \int_{0}^{1-p} t^{n-a-1}(1-t)^a dt
    \label{eq:binom_prob}
\end{align}

 \begin{lemma}
\cite[Eq.~8.17.4]{NIST:DLMF}: For $a,b > 0, \ 0 \leq x \leq 1$,
\begin{align}
    I_x(a,b) = 1 - I_{1-x}(b,a)  \label{eq:I_oneminusx}
\end{align}
 \end{lemma}
 
 \begin{lemma}
\cite[Eq.~8.17.20]{NIST:DLMF}: For $a,b > 0, \ 0 \leq x \leq 1$,
\begin{align}
    I_x(a+1,b) = I_x(a,b) - \frac{x^a(1-x)^b}{a B(a,b)}  \label{eq:I_KandR}
\end{align}
 \end{lemma}


 \begin{lemma}
 The equation $I_{\alpha}(r,r) = c\alpha$ has only one solution when $r>0$, $r \in \mathbb{Z}^+$, $0<\alpha \leq 1/2$ and $0<c\leq 1$. 
 \label{corr:one_soln}
 \end{lemma}

Proof:
First, $\alpha=0$ is a solution to this equation since when $\alpha=0$, both $I_{\alpha}(r,r)$ and  $c\alpha$ are zero. Also, when $c=1$, $\alpha=1/2$ is a solution of the equation since $I_{1/2}(r,r)=1/2$. The derivative of both terms with respect to $\alpha$ is:
\begin{align}
    \frac{\partial (I_{\alpha}(r,r))}{\partial \alpha} &= \frac{\alpha^{r-1}(1-\alpha)^{r-1}}{B(r,r)} , \quad
    \frac{\partial (c \alpha)}{\partial \alpha} = c
\end{align}
It can be seen that the derivative of $c\alpha$ is a constant, and  $\frac{\left(\alpha(1-\alpha)\right)^{r-1}}{B(r,r)}=0$ when $\alpha=0$ and $\frac{\left(\alpha(1-\alpha)\right)^{r-1}}{B(r,r)}$ is monotone increasing in the range $0<\alpha \leq 1/2$. For the case where $c=1$; $\alpha=0$ and $\alpha=1/2$ is a solution to the equation, hence for some $0<\alpha^{**}<1/2$ that satisfies $\frac{\left(\alpha^{**}(1-\alpha^*)\right)^{r-1}}{B(r,r)} = 1$, it must hold that $\frac{\partial (I_{\alpha}(r,r))}{\partial \alpha} < \frac{\partial (\alpha)}{\partial \alpha} =1 $ when $0< \alpha<\alpha^{**}$, and $\frac{\partial (I_{\alpha}(r,r))}{\partial \alpha} > \frac{\partial (\alpha)}{\partial \alpha} =1 $ when $ \alpha^{**} <\alpha \leq 1/2$. This is because if such $\alpha^{**}$ such that $0<\alpha^{**}<1/2$ does not exist, $\alpha=1/2$ can't be a solution to the equation $I_{\alpha}(r,r) = c\alpha$. Now, considering the case for arbitrary $0< c \leq 1$, since $\frac{\alpha^r(1-\alpha)^r}{B(r,r)}=0$ is monotone increasing, there can only be one $0<\alpha^*<1/2$ such that $\frac{(\alpha^*)^r(1-\alpha^*)^r}{B(r,r)} = c$. This means that $c\alpha$ is increasing faster than $I_{\alpha}(r,r)$ in the region $0<\alpha<\alpha^*$, hence there can't be a solution to $I_{\alpha}(r,r)=c\alpha$ in this region. Further, $I_{\alpha}(r,r)$ is increasing faster than $c\alpha$ in the region $\alpha^*<\alpha\leq 1/2$, hence there can be at most one solution to the equation $I_{\alpha}(r,r) = c\alpha$ in the region $0 <\alpha \leq 1/2$. Now, consider the fact that  $I_{\alpha^*}(r,r) < c\alpha^*$, and $I_{1/2}(r,r) = 1/2 \geq c/2$ when $\alpha=1/2$. Combining this with the fact that both functions are continuous, there must be at least one solution to the equation $I_{\alpha}(r,r)=c \alpha$ for $0<c \leq 1$ in the range $\alpha^*<\alpha\leq 1/2$. Combining this with previous statement (that there can be at most one solution in this range), it can be concluded that there is only one solution to the equation $I_{\alpha}(r,r) = c\alpha$ in the range $0<\alpha \leq 1/2$ where $0<c<1$.

\subsection{Proof of Theorem \ref{theorem:thmr_1}}

To prove Theorem \ref{theorem:thmr_1}, we  need to show that   for any $r \in \mathbb{Z}^+$, the random K-out graph $\hhn$ is $r$-robust \emph{whp} if $K_n \geq 2r$. To do this, similar to the proof given in \cite{sundaram_robustness} for Erd\H{o}s-R\'enyi graphs, we will first find an upper bound on the probability of a subset of given size being not $r$-reachable, and then use this result to show that the probability of not being $r$-robust goes to zero when $n \to \infty$ and $K_n \geq 2r$. Different from the prior work which relied on the commonly used  upper bounds for the binomial coefficients ${n \choose k}\leq \left(\frac{en}{k}\right)^k$ and the union bound \cite{sundaram_robustness,can_cdc2021} to  bound  the probability of a subset of given size being not $r$-reachable, our proof  uses the Beta function $B(a,b)$ and its  properties described in the previous Section to achieve tighter bounds. This in turn enables us to establish a tighter threshold for the $r$-robustness of random K-out graphs than what was previously possible; e.g., see \cite{can_cdc2021}.

First, let $\mathcal{E}_n (K_n, r; S)$ denote the event that  $S \subset V$ is an $r$-reachable set as per Definition~\ref{def:r-reachable}. The event $\mathcal{E}_n (K_n, r; S)$ occurs if there exists at least one node in $S$ that is adjacent to at least $r$ nodes in $S^c$, the subset comprised of nodes outside the subset $S$. Thus, we have
\begin{align}
\mathcal{E}_n (K_n, r; S) = 
\bigcup_{i \in \nodes_{S}}  \left \{ \left( \sum_{j \in \nodes_{S^c}} \mathds{1} \left \{ v_i \sim v_j  \right \} \right) \geq r  \right \}  \nonumber
\end{align}
with $\nodes_{S}$,  $\nodes_{S^c}$ denoting the set of labels of the vertices in  $S$ and $S^c$, respectively, and $\mathds{1} \{ \}$ denoting the indicator function.
We are also interested in the complement of this event, denoted as $\left (\mathcal{E}_n (K_n, r; S) \right)\comp$, which occurs if all nodes in $S$ are adjacent to less than r nodes in $S^c$. This can be written as
\begin{align}
\left( \mathcal{E}_n\comp (K_n, r; S) \right)  = 
\bigcap_{i \in \nodes_{S}} \left \{ \left( \sum_{j \in \nodes_{S^c}} \mathds{1} \left \{ v_i \sim v_j  \right \} \right) < r  \right \}.  \nonumber
\end{align}
Note that at least one subset in every disjoint subset pairs needs to be $r$-reachable per the definition of $r$-robustness, hence one of the events $\mathcal{E}_n (K_n, r; S)$ or $\mathcal{E}_n (K_n, r; S')$ need to hold {\em with high probability} for every disjoint subset pairs $S, S'$ of $V$. Now, let $\mathcal{Z}(K_n, r)$ denote the event that both subsets in at least one of the disjoint subset pairs $S, S' \subset V$ are $r$-reachable. Thus, we have  

\begin{align}
\mathcal{Z}(K_n, r) & = \hspace{-4mm} \bigcup_{S, S' \in \mathcal{P}_n: ~  |S| \leq \lfloor \frac{n}{2} \rfloor} \hspace{-3mm} \left[ (\mathcal{E}_n({K}_n,r; S))\comp \wedge (\mathcal{E}_n({K}_n,r; S\comp))\comp \right], \nonumber
\end{align}
where $S \cap S' = \emptyset$, $\mathcal{P}_n$ is the collection of all non-empty subsets of $V$ and since for each $S$ we check the $r$-reachability of both $S$ and $S\comp$, the condition $| S | \leq \lfloor \frac{n}{2} \rfloor$ is used to prevent counting each subset twice. From this defintion, it can be seen that the graph $\hhn$ is $r$-robust if the event $\mathcal{Z}(K_n, r)$ does not occur. Using union bound, we get
\begin{align}
P_Z &\leq \hspace{-3mm}  \sum_{ |S| \leq \lfloor \frac{n}{2} \rfloor } \pr[ \left( \mathcal{E}_n (K_n, r; S)\right) \comp \wedge \left( \mathcal{E}_n (K_n, r; S')\right) \comp ] \nonumber \\
&\leq \hspace{-3mm}  \sum_{ |S| \leq \lfloor \frac{n}{2} \rfloor } \pr[ \left( \mathcal{E}_n (K_n, r; S)\right) \comp   ] \nonumber \\
&=\hspace{-1mm} \sum_{m=1}^{ \left\lfloor \frac{n}{2} \right\rfloor }
 \sum_{S_m \in \mathcal{P}_{n,m} } \pr[\left( \mathcal{E}_n (K_n, r; S_m)\right) \comp  ]  \label{eq:UnionBound},
\end{align}

where $\mathcal{P}_{n,m}$ denotes the collection of all subsets of $V$ with exactly $m$ elements, and let $S_m \in \mathcal{P}_{n,m}$ be a subset of the vertex set $V$ with size $m$, i.e. $S_m \subset V$ and $|S_m| =m$. Further,   $\pr\left[\mathcal{Z}(K_n,r) \right]$ is abbreviated as $P_Z := \pr\left[\mathcal{Z}(K_n,r) \right]$. 
 From the exchangeability of the node labels and associated random variables, we have

\begin{align}
\sum_{S_m \in \mathcal{P}_{n,m} }   \pr[\left( \mathcal{E}_n (K_n, r; S_m)\right) \comp  ]   = {n\choose m} \pr[\left( \mathcal{E}_n (K_n, r; S_m)\right) \comp  ]. 
\label{eq:ForEach=r}
\end{align}

since $|\mathcal{P}_{n,m} | = {n \choose m}$, as there are ${n \choose m}$ subsets of $V$ with $m$ elements. Substituting this into (\ref{eq:UnionBound}), we obtain 
\begin{align}
P_Z \leq \sum_{m=1}^{ \left\lfloor \frac{ n}{2} \right\rfloor } {n\choose m} \pr[\left( \mathcal{E}_n (K_n, r; S_m)\right) \comp  ] \nonumber
\label{eq:Zbound}
\end{align}
\vspace{-2mm}

Before evaluating this expression, we will start with evaluating the probability that the set $S_m$ is not $r$-reachable, abbreviated as $\pr[\left( \mathcal{E}_n (K_n, r; S_m)\right) \comp ] :=  P_{S_m}$. Since a node $v \in S_m$ can have neighbors in $S_m^c$ if it forms an edge with nodes in $S_m^c$ or if nodes in $S_m^c$ forms edges with node $v$, let $P_{S_{m,1}}$ denote the probability that all nodes $v \in S_m$ form an edge with less than $r$ nodes in $S_m^c$, and let $P_{S_{m,2}}$ denote the probability that for each node $v \in S_m$, nodes in $S_m^c$ form less than $r$ edges with them. Evidently, $P_{S_m} \leq P_{S_{m,1}} \cdot P_{S_{m,2}}$. Further, let $P_{v_{m,1}}$ denote the probability that a node $v \in S_m$ forms an edge with less than $r$ nodes in $S_m^c$, and let $P_{v_{m,2}}$ denote the probability that nodes in $S_m^c$ form less than $r$ edges with the node $v \in S_m$.

\begin{lemma}
The probability that the node $v \in S_m$ chooses less than $r$ nodes in the set $S_m^c$, denoted as $P_{v_{m,1}}$, can be upper bounded by the cumulative distribution function $F(r-1;K_n,p)$ of a binomial random variable with $K_n$ trials and success probability $p=\frac{n-m-r+1}{n-r}$. 
\label{corrol:prob1}
\end{lemma}

Proof: For node $v$, after making one selection, the number of nodes available to choose from decreases so the probability of choosing a node in $S_m^c$ changes at each selection. For example, the probability of choosing a node in $S_m^c$ in the first selection is $\frac{n-m}{n-1}$ and the probability of choosing a node in $S_m^c$ in the second selection is $\frac{n-m-1}{n-2}$ if a node in $S_m^c$ was selected in the first selection and it is $\frac{n-m}{n-2}$ otherwise. Based on this, the probability of selecting a node in $S_m^c$ at the $i^{th}$ selection out of $K_n$ selections can be expressed as $\frac{n-m-j}{n-i}$, $1 \leq i \leq K_n$, $0 \leq j < i$ where $j$ denotes the number of nodes already chosen from the set $S_m^c$ before the $i^{th}$ selection. Since we are considering the case of choosing less than $r$ nodes in $S_m^c$, we have that $j < r$, and with this constraint the lowest possible value of $\frac{n-m-j}{n-i}$ occurs when $j=r-1$ and $i=r$, and hence it is $\frac{n-m-r+1}{n-r}$. This gives a lower bound on the probability of selecting a node in $S_m^c$ in one of the $K_n$ selections and at the same time is an upper bound on the probability of not  selecting a node in $S_m^c$ in one of the $K_n$ selections, and hence it is an upper bound for choosing less than $r$ nodes.

Next, using this upper bound, we plug in $n=K_n$ and $p=\frac{n-m-r+1}{n-r}$ to \eqref{eq:binom_prob}, then we have
\begin{align}
    P_{v_{m,1}} & \leq F\left(r-1;m,1- \frac{m-1}{n-r}\right) = I_{\frac{m-1}{n-r}}(K_n-r+1,r) \nonumber \\ & = (K_n -r+1){K_n \choose r-1} \int_{0}^{\frac{m-1}{n-r}} t^{K_n-r}(1-t)^{r-1}dt 
\end{align}

The selections of each node in $S_m$ are independent, hence we can use $(P_{S_{m,1}}) =(P_{v_{m,1}})^m$.

In order to find $P_{v_{m,2}}$, a node in $S_m^c$ forming an edge with the node $v$ can be modeled as a Bernoulli trial with probability $p=\frac{K_n}{n-1}$ so the event that nodes in $S_m^c$ forming less than $r$ edges with the node $v$ can be represented by a Binomial model with $n-m$ trials and $p=\frac{K_n}{n-1}$. Hence,
\begin{align}
    P_{v_{m,2}} & = F\left(r-1;n-m, \frac{K_n}{n-1}\right) = I_{\frac{n-K_n-1}{n-1}}(n-m-r+1,r)
\end{align}


Since the nodes in $S_m^c$ forming edges with nodes in $S_m$ are not independent of the other nodes in $S_m$, we cannot write $(P_{S_{m,2}}) \leq (P_{v_{m,2}})^m$. To find $(P_{S_{m,2}})$, we will decompose it into the following conditional probabilities.
\begin{align}
    P_{S_{m,2}} = & \pr[d_{v_1}<r] \cdot \pr[d_{v_2}<r | d_{v_1}<r] \cdot \nonumber \\ & \ldots \pr[d_{v_m}<r | d_{v_1}<r, d_{v_2}<r, \ldots, d_{v_{m-1}}<r]
\end{align}

where $v_1, v_2, \ldots, v_m  \in S_m$ represent all the nodes in $S_m$, and $d_{v_i}$ is used to denote the number of nodes in $S_m^c$ that form an edge with the node $v_i$. To find an upper bound on $P_{S_{m,2}}$, we consider the worst case. In the worst case, all the preceding nodes are selected by nodes in $S_m^c$ exactly $r-1$ times. This reduces the number of available edges in $S_m^c$ that can make connections with the remaining nodes in $S_m$, hence increases the probability of nodes in $S_m^c$  forming less than $r$ edges with the remaining nodes in $S_m$. Hence, we can write:

\begin{align}
    P_{S_{m,2}} & \leq  \pr[d_{v_1}<r] \cdot \pr[ d_{v_2}<r | d_{v_1}=r-1] \cdot \nonumber \\ & \ldots \pr[d_{v_m}<r | d_{v_1}=r-1,  \ldots, d_{v_{m-1}}=r-1]
\end{align}

Consider the general case for $\pr[d_{v_{a+1}}<r | d_{v_1}=r-1,  \ldots, d_{v_{a}}=r-1]$ where $1\leq a \leq m-1$. Assume that $q_1$ nodes in $S_m^c$ formed an edge with only one node among the nodes $v_1,\ldots,v_a$. Similarly, assume $q_2$ nodes in $S_m^c$ formed an edge with only two nodes, and so on ($q_{K_n}$ nodes in $S_m^c$ formed edges with $K_n$ nodes in $v_1,\ldots,v_a$). Also define $q=q_1 + q_2 + \ldots + q_{K_n}$. It can be seen that $a = \frac{q_1 + 2q_2 + \ldots + K_n q_{K_n}}{r-1}$. Here, we have $n_0 = n-m-q$ nodes that did not use any of its selections yet, so their probability of choosing the node $v_{a}$ is $p_0 = \frac{K_n}{n-a-1}$. Similarly, we have $n_1 = q_1$ nodes that used one of its selections, so their probability is $p_1 = \frac{K_n -1 }{n-a-1}$ (for $n_{K_n}=q_{K_n}$ nodes $p_{K_n}=\frac{K_n - K_n }{n-a-1}=0$).

Considering the selection of each node in $S_m^c$ as a Bernoulli trial with different probabilities, the collection of all such trials defines a Poisson Binomial distribution $X \sim PB([p_0]^{n_0},\ldots,[p_{K_n}]^{n_{K_n}})$. The mean of this distribution is:
\begin{align}
    \mu_X &= (n-m-q)*\frac{K_n}{n-a-1} + \sum_{i=1}^{K_n} q_i\frac{K_n-i}{n-a-1} \nonumber \\
    &= \frac{(n-m)K_n-a(r-1)}{n-a-1}
\end{align}

Assuming $K_n > 2r-2$ and using $m \leq \lfloor n/2 \rfloor$, we have:
\begin{align}
    \mu_X = \frac{(2n-2m-a)}{n-a-1} *(r-1) +   \frac{K_n - 2r + 2}{n-a-1}
\end{align}

Since the mode of the Poisson Binomial distribution satisfies $\psi_X \leq \mu_X + 1$ \cite{poissonbinomial_mode}, we have $\psi_X > r-1$ for any $1\leq a \leq m-1$ if $K_n > 2r-2$. Since for any distribution, the cumulative distribution function is equal to 1/2 at the mode of distribution, we have that $\pr[d_{v_{a+1}} < r | d_{v_1}=r-1,  \ldots, d_{v_{a}}=r-1] < 1/2$. Hence,
\begin{align}
    P_{S_{m,2}} < \left(\frac{1}{2}\right)^m
\end{align}


Now, using the fact that $P_{S_m} \leq P_{S_{m,1}} \cdot P_{S_{m,2}}$, we have:

\begin{align*}
    P_{S_m} & \leq P_{S_{m,1}} \cdot P_{S_{m,2}} \leq (P_{v_{m,1}})^m \cdot \left(\frac{1}{2}\right)^m \nonumber \\ & \leq \hspace{-1mm} \left( \frac{1}{2} I_{\frac{m-1}{n-r}}(K_n \hspace{-0.3mm} - \hspace{-0.3mm} r \hspace{-0.3mm} + \hspace{-0.3mm} 1,r) \hspace{-0.3mm}  \hspace{-0.3mm}\right)^{m} 
\end{align*}



Let $P_m := {n\choose m}  P_{S_m} $. Then, we have:
\begin{align}
    P_Z \leq \sum_{m=1}^{ \left\lfloor \frac{ n}{2} \right\rfloor } {n\choose m} P_{S_m} = \sum_{m=1}^{ \left\lfloor \frac{ n}{2} \right\rfloor } P_m
\end{align}

We will divide the summation into three parts as follows:
\begin{align}
    P_Z = \sum_{m=1}^{\lfloor n/2 \rfloor} P_m = \sum_{m=1}^{\lfloor \log(n) \rfloor} P_m + \sum_{m=\lceil \log(n) \rceil}^{\lfloor \alpha^*(n - r) \rfloor} P_m \nonumber \\ + \sum_{m=\lceil \alpha^*(n - r)  \rceil}^{\lfloor n/2 \rfloor} P_m = P_1 + P_2 + P_3
\end{align}
where $\alpha^*$ is the solution to the equation $I_{\alpha}(r,r) = \frac{n-r}{ne}\cdot \alpha$ in the range $0<\alpha<\frac{1}{2}$. (The purpose of defining $\alpha^*$ this way will be given later in the proof.) 
Start with the first summation $P_1$ and use  \eqref{eq:binomial_property} along with ${n \choose m} \leq \left( \frac{en}{m} \right)^m$, then we have:

\begin{align}
    P_m & = {n \choose m} P_{S_m}  \leq  {n \choose m}  P_{S_{m,1}} \cdot P_{S_{m,2}} \leq {n \choose m}  P_{S_{m,1}}  \nonumber \\
   & \leq  \left(\frac{en}{m} I_{\frac{m-1}{n-r}}(K_n-r+1,r) \right)^{m} 
\nonumber  \\ &
 \leq  \left(\frac{en}{m B(K_n-r+1,r)} \int_{0}^{\frac{m-1}{n-r}} t^{K_n-r}(1-t)^{r-1}dt \right)^{m} 
   \nonumber  \\ &
     \leq  \left(\frac{en}{m B(K_n-r+1,r)} \int_{0}^{\frac{m-1}{n-r}} t^{K_n-r}dt \right)^{m} \nonumber  \\ &
   \leq  \left(\frac{en \left( \frac{m-1}{n-r}\right)^{K_n-r+1}}{m B(K_n-r+1,r) (K_n-r+1)}  \right)^{m}  \nonumber  \\ &  \leq  \left( \frac{e (1+r/m)}{B(K_n-r+1,r)(K_n-r+1)} \left( \frac{m-1}{n-r}\right)^{K_n-r} \right)^{m} \nonumber  \\ &  \leq  \left( \frac{e (1+r) \left( \frac{\log(n)-1}{n-r}\right)^{K_n-r}}{B(K_n-r+1,r)(K_n-r+1)}  \right)^{m} := (a_n)^m \nonumber
\end{align}

For $K_n > r$, since $B(K_n-r+1,r)$ and $r$ are finite values, we have $\limit a_n = 0$ by virtue of $\limit \left(\frac{\log(n)-1}{n-r}\right)^{K_n-r} = 0$.  Using this, we can express the summation as:
\begin{align}
    P_1 = \sum_{m=1}^{\lfloor \log(n) \rfloor} (a_n)^m \leq a_n \cdot \frac{1 - (a_n)^{\log(n)}}{1 - a_n}
\end{align}
where the geometric sum converges by virtue of $\lim _{n \to \infty} a_n = 0$, leading to $P_1$ converging to zero for large $n$. 

Now, similarly consider the second summation $P_2$. Using ${n \choose m} \leq \left( \frac{en}{m} \right)^m$, we have
\begin{align}
    P_m & \leq  {n \choose m} P_{S_m} \leq {n \choose m} (P_{S_{m,1}})^m  \nonumber \\
    &\leq  \left(\frac{en}{m} I_{\frac{m-1}{n-r}}(K_n-r+1,r) \right)^{m} := (a_m)^m
\end{align}
where $a_m := \frac{en}{m} \cdot  I_{\frac{m-1}{n-r}}(K_n-r+1,r)$. Assume that $K_n > 2r-1$. It can be shown that $I_{\frac{m+r}{n}}(K_n -r+1) < I_{\frac{m+r}{n}}(r,r)$ as a consequence of the property (\ref{eq:I_KandR}). Hence, we have
\begin{align}
    a_m & <  \frac{en}{m}\cdot I_{\frac{m-1}{n-r}}(r,r) \leq \frac{en}{m}\cdot I_{\frac{m}{n-r}}(r,r)
\end{align}

Assume that $\alpha=\alpha^*$ is the solution of the equation $I_{\alpha}(r,r) = \frac{n-r}{ne}\cdot \alpha$ in the range $0<\alpha<\frac{1}{2}$. From Lemma \ref{corr:one_soln}, we know that this equation has only one solution in this range, and $I_{\alpha}(r,r) \leq \frac{\alpha^* (n-r)}{ne}$ for $0<\alpha \leq \alpha^*$. Plugging in $\alpha = \frac{m}{n-r}$, we have $I_{\frac{m}{n-r}}(r,r) \leq \frac{\alpha^* m}{ne}$, which leads to $a_m < 1$ when $m \leq \lfloor \alpha^*(n-r) \rfloor$. Denoting $a:= \max_m (a_m)$, we have 
\begin{align}
    P_2 =   \sum_{m=\lceil \log(n) \rceil}^{\lfloor \alpha^*(n - r) \rfloor} (a_m)^m \leq \sum_{m=\lceil \log(n) \rceil}^{\infty} (a)^m \leq \frac{a^{\log(n)}}{1-a}
\end{align}
where the geometric sum converges by virtue of $\lim _{n \to \infty} a < 1$, leading to $P_2$ converging to zero as $n$ gets large.

Now, similarly consider the third summation $P_3$. Assuming $K_n \geq 2r-1$, we have

\begin{align}
    P_m & \leq  {n \choose m} (P_{S_{m,1}})^m \cdot (P_{S_{m,2}})^m  \nonumber \\
    & <  \left(\frac{1}{2} {n \choose m}^{\frac{1}{m}} I_{\frac{m-1}{n-r}}(K_n-r+1,r) \right)^{m} \nonumber \\
    & <  \left(\frac{1}{2} {n \choose m}^{\frac{1}{m}} I_{\frac{m}{n-r}}(r,r) \right)^{m} := (a_m)^m   
\end{align}

From the previous summation $P_2$ and the proof of Lemma \ref{corr:one_soln}, we know that in the range $\lceil \alpha^*n - r  \rceil \geq m \geq \lfloor n/2 \rfloor$; $\frac{n}{m}$ is decreasing, $I_{\frac{m}{n-r}}(r,r)$ is increasing, and the overall expression $\frac{n}{m} \cdot I_{\frac{m}{n-r}}(r,r)$ is also increasing. Hence, $\frac{n}{m}\cdot I_{\frac{m}{n-r}}(r,r)$ takes its maximum value when $m=\lfloor n/2 \rfloor$. Since $\frac{en}{m}$ is an upper bound of ${n \choose m}^{\frac{1}{m}}$, the expression $a_m$ also takes its maximum value when $m=\lfloor n/2 \rfloor$. Denoting $a:= \max_m (a_m) = a_{\lfloor n/2 \rfloor}$, and using Stirling's approximation $\limit {2n \choose n} \sim \frac{4^n}{\sqrt{\pi n}} < \frac{4^n}{\sqrt{\pi}}$ as a finer upper bound, we have 

\begin{align}
    \limit a = \limit a_{\lfloor n/2 \rfloor} <  \frac{1}{2} \cdot \frac{4}{\sqrt{\pi} } \cdot \frac{1}{2} < 1 
\end{align}

From this, we have:

\begin{align}
     P_3 & \leq  \sum_{m=\lceil \alpha^*(n - r)  \rceil}^{\lfloor \frac{n}{2} \rfloor} (a_m)^n \leq \frac{n}{2} (a)^n
\end{align}

where the sum converges by virtue of $\limit a = 1$ since $\limit n(a)^n = 0 $ in this case, leading to $P_3$ converging to zero as $n$ gets large. Since $P_1$, $P_2$, and $P_3$ all converge to zero as $n$ gets large, $P_Z = P_1 + P_2 + P_3$ also converges to zero as $n$ gets large. This concludes the proof of Theorem \ref{theorem:thmr_1}.

\section{Conclusion} \label{sec:conclusion}

In this paper, we provide a comprehensive set of results on the $r$-robustness of the random K-out graph $\hhn$, and the connectivity and giant component size of $\hhdn$, i.e., random K-out graph with (randomly selected) $\gamma_n$ nodes deleted. In addition to providing proofs of our results, we include computer simulations to validate our results in the finite node regime. To demonstrate the usefulness of the random K-out graphs, we compare our results on the random K-out graphs with results from ER graphs under similar settings, and determine that random K-out graphs attain $r$-robustness, connectivity or the occurrence of a giant component of a given size at a significantly lower mean node degree value compared to ER graphs. These results reinforce the usefulness of random K-out graphs in applications that require a certain degree of robustness or tolerance to nodes failing, being captured, or being dishonest; such as federated learning, consensus dynamics, distributed averaging and wireless sensor networks.

\appendices

\bibliographystyle{IEEEtran}
\bibliography{IEEEabrv,references}

\end{document}